\documentclass[sigconf]{acmart}
\usepackage[utf8]{inputenc}
\usepackage[T1]{fontenc}

\usepackage{amsmath}
\usepackage{algorithm}
\usepackage[labelfont=bf,textfont=md]{caption}
\usepackage[noend]{algpseudocode}
\usepackage{booktabs}
\usepackage{multirow}
\usepackage{caption}
\usepackage{subcaption}
\usepackage{float}
\usepackage{enumitem}
\usepackage{tikz}
\usepackage{lipsum}

\newcommand\blfootnote[1]{%
  \begingroup
  \renewcommand\thefootnote{}\footnote{#1}%
  \addtocounter{footnote}{-1}%
  \endgroup
}

\copyrightyear{2022}
\acmYear{2022}
\setcopyright{acmcopyright}
\acmConference[SIGMOD '22]{Proceedings of the 2022 International Conference on Management of Data}{June 12--17, 2022}{Philadelphia, PA, USA}
\acmBooktitle{Proceedings of the 2022 International Conference on Management of Data (SIGMOD '22), June 12--17, 2022, Philadelphia, PA, USA}
\acmPrice{15.00}
\acmDOI{10.1145/3514221.3526167}
\acmISBN{978-1-4503-9249-5/22/06}

\title{Proteus: A Self-Designing Range Filter}

\author{Eric R. Knorr*\textsuperscript{1}, Baptiste Lemaire*\textsuperscript{1}, Andrew Lim*\textsuperscript{1} \\ Siqiang Luo\textsuperscript{2}, Huanchen Zhang\textsuperscript{3}, Stratos Idreos\textsuperscript{1}, Michael Mitzenmacher\textsuperscript{1}}
\affiliation{%
 \institution{\textsuperscript{1}Harvard University, \textsuperscript{2}Nanyang Technological University, \textsuperscript{3}Tsinghua University}
}

\begin{abstract}
We introduce Proteus, a novel self-designing approximate range filter, which configures itself based on sampled data in order to optimize its false positive rate (FPR) for a given space requirement.
Proteus unifies the probabilistic and deterministic design spaces of state-of-the-art range filters to achieve robust performance across a larger variety of use cases.
At the core of Proteus lies our Contextual Prefix FPR (CPFPR) model \textemdash a formal framework for the FPR of prefix-based filters across their design spaces.
We empirically demonstrate the accuracy of our model and Proteus' ability to optimize over both synthetic workloads and real-world datasets.
We further evaluate Proteus in RocksDB and show that it is able to improve end-to-end performance by as much as 5.3x over more brittle state-of-the-art methods such as SuRF and Rosetta.
Our experiments also indicate that the cost of modeling is not significant compared to the end-to-end performance gains and that Proteus is robust to workload shifts.
\blfootnote{*These authors contributed equally.\vspace{-0.5em}}
\end{abstract}



\begin{document}

\maketitle

\setlength{\belowcaptionskip}{-8pt}

\section{Introduction}
\label{sec:intro}

\begin{figure}
    \centering
    \includegraphics[trim= 2 0 2 0, clip=true, width=0.46\textwidth]{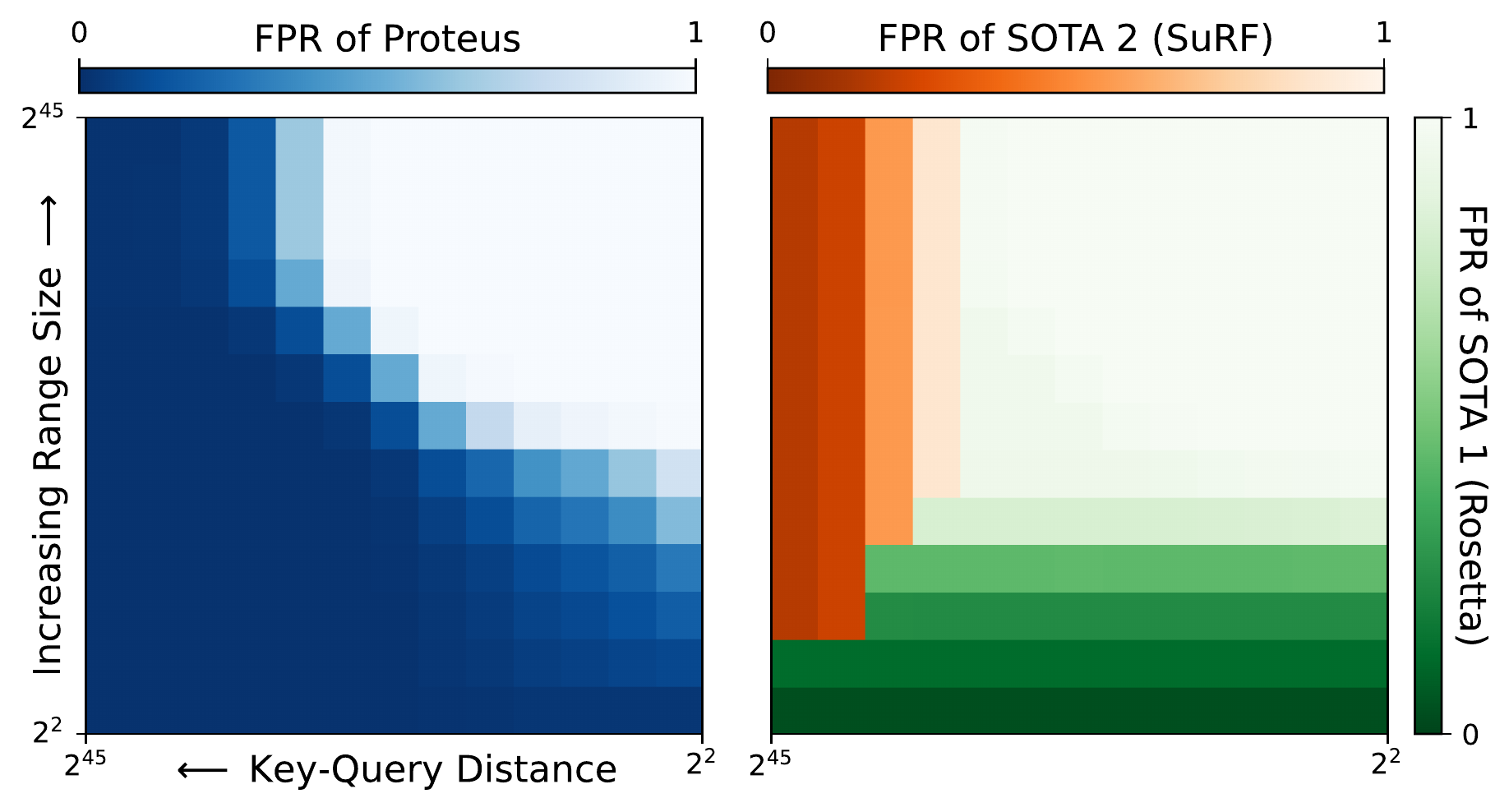}
    \vspace{-1em}
    \caption{A self-designing filter achieves superior performance in a wide variety of workloads. (Darker is better, lower FPR.)}
    \vspace{-1.4em}
    \label{fig:teaser}
\end{figure}

\noindent \textbf{The Importance of Range Filters:}
Range queries are a fundamental operation in big data applications.
Given a set $S$, a range query $\texttt{[a,b]}$ returns the members of $S$ within the query interval, i.e. $S \cap \texttt{[a,b]}$.
Example applications that need range queries and handle large amounts of data include social media platforms using spatio-temporal queries to aggregate user events \cite{gisdatabase}, pattern discovery and anomaly detection in time-series data streams \cite{kondylakis2020coconut}, scientific spatial models \cite{spatialrangequery}, graph databases \cite{dgraph, kyrola2014graphchidb} and Blockchain analytics \cite{vchain}. 
Range queries over such data sets are expensive due to the disk or network costs required to process the data. 
Using a filter data structure to determine when no elements are in the query range can vastly improve performance by preventing unnecessary IO operations.

As an important unifying application, large-scale data systems keep large volumes of data on cheap but high latency storage devices. 
Answering range queries requires checking every data page that intersects with the queried range to retrieve the relevant data. 
For example, in widely used Log Structured Merge (LSM) tree-based key-value stores, one or more pages must be checked per level \cite{sivasubramanian2012amazon, alsubaiee2014asterixdb, chang2008bigtable, dong2021evolution, lakshman2010cassandra}.
However, explicitly reading in all intersecting data pages would be far too expensive. 
In this setting, a compact range query data structure can be set up on a per page basis.
Range query structures then act as filters, which can determine that no data in a range exists on a page (that is, we have an empty query) before resorting to accessing disk. 
As such, much of our evaluation considers the effectiveness of  range filters in LSM-trees.

\noindent \textbf{Strong Guarantees are Impractical:} 
Most work on filtering has been focused on single key queries. Such filters are referred to as Approximate Membership Query structures (AMQs).
AMQs such as Bloom filters \cite{bloom_filter} and their many variants are used to avoid the majority of unnecessary lookups for individual items in diverse applications.
AMQs tolerate a small probability of false positives in order to achieve a compact representation of the key set that supports membership queries. 
The false positive rate (FPR) is the primary metric of a filter's performance, as false positives incur unnecessary lookups to verify their emptiness.
Similar techniques can be applied to the approximate range emptiness problem to avoid unnecessary range lookups, but guaranteeing a low FPR for all potential queries can be expensive.  
Prior work has shown that to guarantee an FPR $\epsilon$ for range emptiness queries of range size $R$ requires $\Omega\left(\log_2(R)+\log_2(1/\epsilon)\right)$ bits per key (BPK) \cite{optimal_space}.
Achieving a small FPR for all queries including for large ranges therefore requires an undesirable memory budget.

\noindent\textbf{Design Tradeoffs are Necessary For Performance:}
Current range filters aim to use less memory than theoretical worst-case guarantees require by {\em not} providing false positive guarantees for all queries.
Instead, state-of-the-art solutions prioritize particular types of workloads, such as those with large ranges \cite{surf} or queries that are correlated to the key set \cite{rosetta}, by making hard-coded design decisions that cater to their target use case.
These range filters use heuristic methods where the FPR depends on the relationship between the key and query sets, and provable guarantees are weak or restricted to specific situations. 
These decisions limit their usefulness as deviation from the intended use cases results in increasingly sub-optimal performance, as seen in Figure \ref{fig:teaser}.
This is an issue for use cases not covered by current range filters, for example, particle physics workloads that contain long ranges and correlated queries when cross referencing time series data from multiple sensors to identify events of interest \cite{amrouche2021tracking}.
Additionally, deviations can also arise from workloads that shift over time.  
For instance, applications with different data based on language, such as Wikipedia, exhibit temporal skew in query distribution due to the correlation between time zone and language \cite{skewed_range_query, skew_aware_db_partition}.
Therefore, a robust range filter requires the ability to prioritize the desired use case by navigating the range filter design space.
However, to our knowledge, there has been no prior work on formalizing the parameters of the range filter design space and the tradeoffs therein.

\noindent \textbf{A Formal Framework Allows for Informed Designs:}
The standard metric for FPR analysis is worst-case performance which fails to capture the nuances of range filter design choices because they optimize for specific use cases.  
We introduce the Contextual Prefix FPR (CPFPR) model which formalizes 
how FPR varies across the design space of prefix-based filters.
We focus on the prefix filter design space as this encompasses all state-of-the-art range filter designs \cite{rocksdb, rosetta, siberia, surf}.
The nuances of each design are captured by expressing their FPR in terms of use case features, such as query range size and the proximity between keys and queries.
These features can be concisely described by the characteristics of shared prefixes and the number of unique prefixes of a given length.
We then apply the CPFPR model to techniques used by state-of-the-art range filters in order to understand their design tradeoffs. 

\noindent \textbf{Self-Designing Approximate Range Filters:}
We use insights from the CPFPR model to develop a novel class of readily optimized range filters, that we have dubbed Protean Range Filters (PRFs).
PRFs navigate some portion of the range filter design space and aim to choose the best design available for a given use case. 
We introduce a novel PRF filter, Proteus, that spans a large portion of the prefix-filter design space and makes use of the CPFPR model to navigate it.
Not only does Proteus encompass the design spaces of state-of-the-art range filters, but it is also able to combine their designs in a complementary fashion for even greater effect.
The user need only supply a sample of example queries to be fed into the CPFPR model. 
Increased design flexibility paired with automated optimization allows Proteus to outperform more constrained designs in the vast majority cases, even when said designs are optimally tuned.
We can see in Figure \ref{fig:teaser} that Proteus achieves a low FPR on a larger region of the workload space as compared to the state-of-the-art range filters which are only optimal within confined, mostly disjoint regions.

\noindent \textbf{Contributions:} Our contributions are as follows:
\begin{itemize}[leftmargin=*]
    \item \textit{Novel Range Filter:} We present a novel range filter that can instantiate configurations from across the current range filter design space to meet the needs of various workloads within a limited memory budget.   
    \item \textit{Formalization of the Range Filter Design Space:} We introduce the CPFPR model which captures the tradeoffs in the design space of prefix-based range filters. We use this model to break down and incorporate the design elements of state-of-the-art range filters into our novel range filter \textbf{(Section \ref{sec:modeling})}.
    \item \textit{Efficiently Leveraging Context:} We demonstrate how to practically navigate the prefix filter design space \textbf{(Section \ref{sec:PRFs})}.
    \item \textit{Model Validation:} We validate the accuracy of our model and demonstrate its ability to optimize across the prefix filter design space in a wide variety of settings using both synthetic workloads and real world datasets \textbf{(Section \ref{sec:standalone})}.
    \item \textit{Robust End-to-End Gains:} We show how PRFs can be integrated in a real world system, RocksDB, and demonstrate PRFs' robustness to changing workloads as well as end to end latency improvements of up to 3.9x on 64-bit integers \textbf{(Section \ref{sec:rocksdb})} and 5.3x on strings \textbf{(Section \ref{sec:strings})}.
\end{itemize}

\section{The Range Filter Design Space}
\label{sec:background}
Current solutions to the approximate range emptiness problem, or Approximate Range Emptiness structures (AREs), use one of two fundamental design elements: probabilistic Approximate Membership Query data structures (AMQs) and deterministic search trees.   
Both strategies use prefixes of the key set to encode the key space at different granularities.  

AMQs are natural building blocks as they are designed to reduce unnecessary lookups by using a compact representation of the data set that is small enough to fit in memory and supports fast membership queries.
A low non-zero false positive probability is tolerated in order for the representation to be sufficiently compact, but never a false negative.
The search tree approach is more novel and explicitly encodes prefixes of the key set as a trie.

\subsection{Probabilistic Prefix Filters}
One method used by current AREs is to include one or more AMQs that encode regions of the key space with a single hashed value.
There are many examples of AMQs, such as Bloom Filters, Cuckoo Filters, Quotient Filters, Xor Filters and Ribbon Filters \cite{bloom_filter, cuckoo_filter, quotient_filter, xor_filter, ribbon_filter}, which we will collectively refer to as Bloom filters.  
Though their specifics vary, Bloom filters generally make use of one or more hash functions to encode the key set as a compact array. 
Hashing allows for fast individual item lookups with a low probability of false positives; however, valuable ordering information is lost in the hashing process. 
In order to know whether a range \texttt{[}$l,r$\texttt{]} contains members of the key set, every key from $l$ to $r$ would need to be queried individually and the probability that at least one of them results in a false positive is then proportional to $r-l$. 
As such, the usefulness of Bloom filters for such queries rapidly declines with the size of the range being queried. 

This can be compensated for by hashing prefixes of the key set rather than each individual key. 
Hashing a given prefix encodes that there is at least one member of the key set with the given prefix.  
This retains the benefits of hashing while allowing queries to the Bloom filter to rule out entire regions of the key space at a time.  

\noindent \textbf{Prefix Bloom Filters:}
Prefix Bloom filters have been in use for some time now, particularly in the context of network routing \cite{prefix_matching}.
By hashing a key prefix of length $l$ bits, the prefix Bloom filter encodes a region of the key space of size $2^{k-l}$, where $k$ is the maximum key length.
Using this strategy, a range of the key space can be queried by querying the Bloom filter for each region that overlaps with the desired range.  
This is well suited to situations with clustered target ranges, like IP addresses in network routing, as the prefix length used can be tuned to the cluster sizes.  
Prefix Bloom filters are also used in key-value stores, but the constraint of encoding the key space as fixed size, prefix defined regions limits their usefulness when the target ranges are not well known.
In particular, prefix Bloom filters are very sensitive to the choice of prefix length which we will discuss in more detail in Section \ref{sec:modeling}.

\noindent \textbf{Rosetta:}
Rosetta \cite{rosetta} is an ARE aimed specifically at range queries in database systems using LSM trees, such as RocksDB \cite{rocksdb}. 
These systems have used prefix Bloom filters with shorter prefixes to filter out large range queries; however, if they are not properly tuned, these prefix filters perform very poorly on small to medium range queries as well as individual key queries.
This becomes even worse if the queries are correlated to the key set and empty queries tend to fall close to the key set.

Conceptually, Rosetta encodes the nodes of an implicit segment tree, or binary trie, representing the entire key space. 
All nodes of a given depth present in the key set are encoded by hashing the corresponding prefixes into a single Bloom filter.
The prefix filter encoding the leaf nodes of the tree is then equivalent to a Bloom filter populated with full key hashes.  
A range is queried by checking for the presence of each node in the sub-tree corresponding to the that range in depth first order.
If any node is not present in its respective prefix filter, the entire sub-tree rooted at that node is known to be empty and is not queried further.  
If a given node may be present, the sub-tree continues to be queried in depth-first order until either a leaf node is reported as present, resulting in a positive range query, or the entire sub-tree is found to be empty. 

In practice, Rosetta does not encode every level of this tree and is configured by apportioning the total memory budget between the prefix Bloom filters encoding each prefix length. 
In particular, Rosetta typically allocates all of its memory budget to the last few prefix lengths.  
This allocation strategy works well for small ranges and point queries, regardless of proximity to the key set,
but larger range queries will still require many Bloom filter queries to cover the query range and performance trends towards that of an AMQ. 
\vspace{-1em}
\subsection{Deterministic Prefix Filters }
Other solutions forgo the benefits of hashing to retain as much ordering information as possible.
These AREs typically make use of an explicitly encoded search tree which performs a similar role to the implicit tree encoded by Rosetta.  
In particular, tries are a well studied method of encoding prefixes in a succinct search tree \cite{trie_memory, trees_higher_degree, succinct_trees}.  
In order to fit these trees in memory, they must be pruned down to prefixes of the key set.
As before, each prefix found in the tree represents a range that has at least one member of the key set, while any prefix not found in the tree will not be present in the key set.  
The pruning therefore introduces the possibility of false positives due to common prefixes between keys and non-keys.

These trees are more flexible than prefix Bloom filters in certain aspects as they can easily encode prefixes of different lengths within the same structure and all sub-prefixes are also retained. 
This comes with the downside of having to pay more memory for longer prefixes, since the entire prefix must be explicitly stored, as opposed to prefix Bloom filters which can dedicate  available memory to any individual prefix length.

\noindent \textbf{SuRF:}
The Succinct Range Filter (SuRF) \cite{surf} is a state-of-the-art ARE that encodes prefixes of the key set as a Fast Succinct Trie (FST). 
It is a static filter and does not adapt to queries. 
The FST in SuRF combines the LOUDS-Dense and LOUDS-Sparse representations from \cite{efficient_trees} to achieve an efficient trie encoding (LOUDS-DS) which allows SuRF to encode longer prefixes than other trie-based prefix filters on the same memory budget.
This encoding can also be searched in constant time, so the query time is independent of the size of the range.  
Despite this, encoding every full key in the FST is still typically too expensive, so SuRF's base configuration prunes the branch for each key to the minimum length prefix that uniquely identifies it in the key set. 
If there is additional memory available it can be used to extend these prefixes. 
It can also be used to store hashes of the key suffixes to help rule out individual keys, though these do not provide any additional benefit for range queries.  

Because SuRF is configured purely based on the key distribution, sparsely populated regions of the key space are encoded more coarsely. 
As with other prefix filters, these coarsely encoded regions are not well suited for filtering queries that are close to the key set.
The requirement that each prefix be at least long enough to be uniquely identified in the key set also means that SuRFs minimum memory usage is determined by the distribution of the keys. 
Despite the encoding being very compact, this can still pose an issue in situations where tight memory budgets must be strictly maintained.

\section{A Contextual Prefix FPR Model}
\label{sec:modeling}

In this section we will formalize how different aspects of a workload affect the performance of prefix-based range filters in order to understand the tradeoffs of different designs.
We use this framework to break down the fundamental components of state-of-the-art AREs and realize a unified design space.

\subsection{The Importance of Prefix Lengths}

We begin by modeling a standard prefix Bloom filter as this forms the basis for our more complex structures.
When considering a single prefix Bloom filter, the only parameter to configure is the choice of prefix length.  
Figure \ref{fig:encoding} shows how a set of 4-bit keys can be encoded using 1-4 bit prefixes.    
Each prefix hashed into the filter encodes that at least one member of the key set contains that prefix; therefore, a short prefix encodes many values while a long prefix may only encode a few. 

Consider a key space $\mathcal{K}$ with total order $\preceq$ \footnote{The total order $\preceq$ depends on the types of keys being used. For instance, integer keys would use the standard less-than-or-equal relation, while string keys would use lexicographical ordering.} and a prefix Bloom filter encoding a key set $K\in \mathcal{K}$ with prefix length $l$ and an FPR of $p$. We use the following terms and notation:

\begin{figure}
    \centering
    \includegraphics[width=0.9\linewidth]{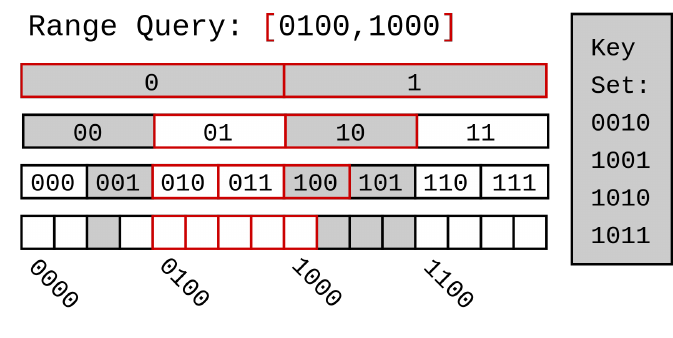}
    \vspace{-1em}
    \caption{Encodings of a 4-bit key space using 1-4 bit prefixes and the respective prefixes required to cover the range \texttt{[4,8]} (outlined in red). Regions containing at least one key are shaded grey.}
    \label{fig:encoding}
    \vspace{0.1em}
\end{figure}

\begin{itemize}
    \item[$Q$:] An empty range query, i.e. an interval of $\mathcal{K}$ s.t. $Q \cap K = \emptyset$. 
    \item[$X_i$:] For an arbitrary set of values $X$ and integer $i$, $X_i$ is the set of unique prefixes $y$ s.t. $len(y) = i$ and $y$ is a prefix of some $x \in X$. Figure \ref{fig:encoding} shows $Q_l$ bordered in red for $Q = \texttt{[0100,1000]}$ and each $l \in \{1,2,3,4\}$.
    \item[LCP:] The longest common prefix. For two arbitrary sets, $X,Y$, we define $lcp(X,Y)$ as the longest LCP between any pair $\{x,y | x\in X, y\in Y\}$.
\end{itemize}

As described in Section \ref{sec:background}, a prefix Bloom filter returns negative only if each prefix in $Q_l$ returns negative.
If $Q_l \cap K_l$ is non-empty, 
then $Q$ is guaranteed to result in a false positive because it contains a prefix used to encode $K$. 
Whether this happens is dependent on the proximity of $Q$ and $K$ which we can express in terms of $lcp(Q,K)$.
If $lcp(Q,K)$ is greater than the prefix length used to encode the key set, then the corresponding members of the key and query set will be indistinguishable from each other.  
This can be seen in Figure \ref{fig:encoding} where only the 4 bit encoding of the key space can distinguish \texttt{1000} from \texttt{1001} as they only differ in the 4th bit. 
Putting all of this together, we can express the probability of a range query false positive in terms of the prefix filter's point query FPR, the range size and proximity of the query to any key, as shown in Equation \ref{eq:pbf}.

\begin{equation}
\label{eq:pbf}
    P_{fp}(Q) = \begin{cases} 
    1-\left(1-p\right)^{|Q_l|}, & \mbox{$lcp(Q,K) < l$} \\
    1, & \mbox{$l \leq lcp(Q,K) $ }
    \end{cases}
\end{equation}

With only a single prefix Bloom filter, there is a clear contention between how many regions must be queried (a short $l$), and whether queries close to the key set can be distinguished from it (a long $l$).    
One option to address this would be to split the memory budget between multiple Bloom filters. 
The downside of this approach is that each Bloom filter may only be able to contribute to a portion of the queries. 
This can be justified if there is a substantial divergence in types of queries such that a single prefix length will not be able to  handle all types of queries effectively.
Take for instance a bimodal distribution of small queries in close proximity to the key set and large non-key-correlated queries.
A single filter tuned to either half of the query set would be effectively useless for the other half.  
However, this requires orders of magnitude of difference between range sizes or key-query correlations for even two filters to be justified as a less performant filter that can contribute to all or most of the queries will result in better overall performance.  
If the distribution is split between additional modes, the possible difference between each decreases and the benefit of further subdividing our memory becomes increasingly marginal.  
As such, we will only consider up to one additional Bloom filter.

Consider then the same setting as before but with two Bloom filters for prefix lengths $l_1 < l_2$ with point query FPRs $p_1$ and $p_2$ respectively.  
We will have to consider several cases depending on how the regions encoded by the first Bloom filter align with $Q$, for which we will use the following additional terms and notation: 
\begin{itemize}
    
    \item[$I_0, I_1$:] Indicator variables for whether the ranges defined by the first and last members of $Q_{l_1}$ are \textit{not} fully contained in $Q$. 
    \item[$I_2, I_3$:] Indicator variables for whether the first and last members of $Q_{l_1}$ are \textit{not} in $K_{l_2}$. If $|Q_{l_1}| = 1$ and $Q_{l_1} \subseteq K_{l_1}$, let $I_2=1$ and $I_3=0$. Note that we cannot have $I_0 = I_2 = 0$ or $I_1 = I_3 = 0$ since $Q$ is empty.
    \item[$L,R$:] The sets of $l_2$ prefixes that intersect with $Q$ and contain the first and last prefixes of $Q_{l_1}$ respectively.
    \item[$P_{l_1}(i)$:] The probability that $i$ of the $l_1$ prefixes completely within $Q$ return false positives.  This can be expressed as the probability mass function for the binomial coefficient with $n=|Q_{l_1}|-I_0-I_1$ and $p = p_1$ as shown in Equation \ref{eq:binom}.
    \item[$\bar{p_L}, \bar{p_R}$:] The probability that $L$ or $R$ respectively are not resolved at the first Bloom filter and return only negatives at the second, given by Equation \ref{eq:LR}. 
\end{itemize}
We now determine the probability of a false positive using two Bloom filters. 
As before, if $l_1 < l_2 \leq lcp(Q,K)$, then $Q$ is guaranteed to result in a false positive.
We therefore consider the case when $lcp(Q,K) < l_2$.
Since $Q \cap K$ is empty, any prefix of $Q_{l_1}$ such that the corresponding set of values is fully contained in $Q$ either yields a false positive or a true negative.  
When a false positive for such a prefix occurs, this yields $2^{l_2 -l_1}$ queries that need to be done at the second Bloom filter. 
However, $Q_{l_1}$ can also share at most two common prefixes with $K$, those at either end of the query;  therefore, these end prefixes may yield false positives or true positives.
Also, these end prefixes, being the only ones that may not be completely within $Q$, result in a number of queries at the second Bloom filter that depends on the overlap between $Q$ and the prefix whenever a positive (false or true) is returned.   
  
The end prefix cases are covered by the $\bar{p_L}$ and $\bar{p_R}$ terms defined in Equation \ref{eq:LR}. 
We then must also sum over the remaining possible combinations of false positives resulting from $Q_{l_1}$. 
Here Equation \ref{eq:binom} gives the probability $P_{l_1}(i)$ of having $i$ false positives that each result in  $2^{l_2-l_1}$ queries at $l_2$.
For each of these we use the complement of the probability that all $l_2$ queries return negative.
Combining these we obtain Equation \ref{eq:dpbf}.

\begin{equation}
    \bar{p_L} = p_1^{I_2}\cdot I_0 \left( 1 - p_2\right)^{|L|}, \\
    \bar{p_R} = p_1^{I_3}\cdot I_1 \left( 1 - p_2\right)^{|R|}
    \label{eq:LR}
\end{equation}

\vspace{-0.1in}

\begin{equation}
    P_{l_1}(i) = \begin{pmatrix} |Q_{l_1}|-I_0-I_1 \\ i  \end{pmatrix} p_1^i(1-p_1)^{|Q_{l_1}|-I_0-I_1-i}
    \label{eq:binom}
\end{equation}

\vspace{-0.1in}

\begin{equation}
    P_{fp}(Q) = 1 -\bar{p_L} -\bar{p_R} -\sum\limits_{i=0}^{|Q_{l_1}|-I_0-I_1} P_{l_1}(i) \left(\left(1-p_2\right)^{i2^{l_2-l_1}} \right)
    \label{eq:dpbf}
\end{equation}

\subsection{Tractable Tries}
By using two Bloom filters with different prefix lengths, one can reasonably address divergent query workloads, but their probabilistic nature can still pose issues.
As discussed before, the longer prefix length is not well suited for large ranges, and any $l_1$ prefix fully within $Q$ that results in a false positive will require $2^{l_2 - l_1}$ prefixes queries at $l_2$. 
If there is much difference between $l_1$ and $l_2$, which is likely if two Bloom filters are justified, then the second Bloom filter will have a very low probability of catching larger ranges missed by the first.  
We could reduce the number of false positives at the first Bloom filter by allocating it a larger proportion of the memory, but this also reduces the filter's ability to deal with correlated queries since this memory must be taken from the second Bloom filter.

Alternatively, using a uniform depth trie at $l_1$ puts a hard limit on the number of prefix queries that may be required at $l_2$ for any range query.  
Consider then the same situation as before but $l_1$ now represents the prefix length of the trie and $p$ is the FPR of the Bloom filter. 
Since the trie is deterministic, only prefixes that match the leaf nodes will ever make it to the Bloom filter.
As discussed prior, only the first and last members of $Q_{l_1}$ can ever match the key set, so no more than $2^{l_2-l_1 +1}-2$ prefixes will need to be queried at $l_2$ for a given range query.  
The probability of a false positive is then given by Equation \ref{eq:hybrid}.

\begin{equation}
\label{eq:hybrid}
    P_{fp}(Q) = \begin{cases} 
    0, & \mbox{$lcp(Q,K) < l_1 < l_2$} \\
    1-\left(1-p\right)^{I_2|L| + I_3|R|}, & \mbox{$l_1 \leq lcp(Q,K) < l_2$} \\
    1, & \mbox{$l_1 < l_2 \leq lcp(Q,K)$ }
    \end{cases}
\end{equation}

Not only is Equation \ref{eq:hybrid} simpler to compute than Equation \ref{eq:dpbf}, but it will achieve a better FPR for any combination of $l_1$ and $l_2$ assuming that the $l_2$ Bloom filter receives the same amount of memory in each.  
This does come with the limitations that $l_1$ now has a fixed memory cost for each possible length.
There is then a hard limit on how long $l_1$ can be and the longer it is, the less memory is available for the $l_2$ Bloom filter.  
Despite this, a short $l_1$ is often cheaper to store explicitly as a trie when compared to the memory a Bloom filter would require to perform comparably.
Tries are particularly efficient when representing clustered data as there will be fewer unique prefixes, but even sparse data sets are relatively cheap to represent as an FST when using a sufficiently short prefix length.

\section{Protean Range Filters}
\label{sec:PRFs}

\noindent \textit{Protean} (pro$\cdot$te$\cdot$an) adj. \textemdash having a varied nature or ability to assume different forms \cite{mw:protean}

\medskip

We define a Protean Range Filter (PRF) as a filter that supports approximate range emptiness queries and configures its own design to optimize performance for any given use case. 
We have presented the CPFPR models for three PRFs: 1PBF \textemdash a standard prefix Bloom filter (Equation \ref{eq:pbf}), 2PBF \textemdash a pair of prefix Bloom filters (Equation \ref{eq:dpbf}) and Proteus \textemdash a hybrid filter that uses both a trie and a prefix Bloom filter (Equation \ref{eq:hybrid}). 
Other than the use of the respective CPFPR models, 1PBF operates as described in Section \ref{sec:background} while 2PBF is equivalent to an instance of Rosetta that uses only 2 filters.
As such, this section will focus primarily on the structure of Proteus and its respective model. 
The models for 1PBF and 2PBF are implemented in a similar fashion.  
We also provide a breakdown of the costs associated with using the model and how these compare for each.
We assume fixed length, integer keys in this section and discuss variable length keys in Section \ref{sec:strings}.

\subsection{Hybrid Architecture}
\label{sec:architecture}

\begin{figure}
    \centering
    \includegraphics[width=\linewidth]{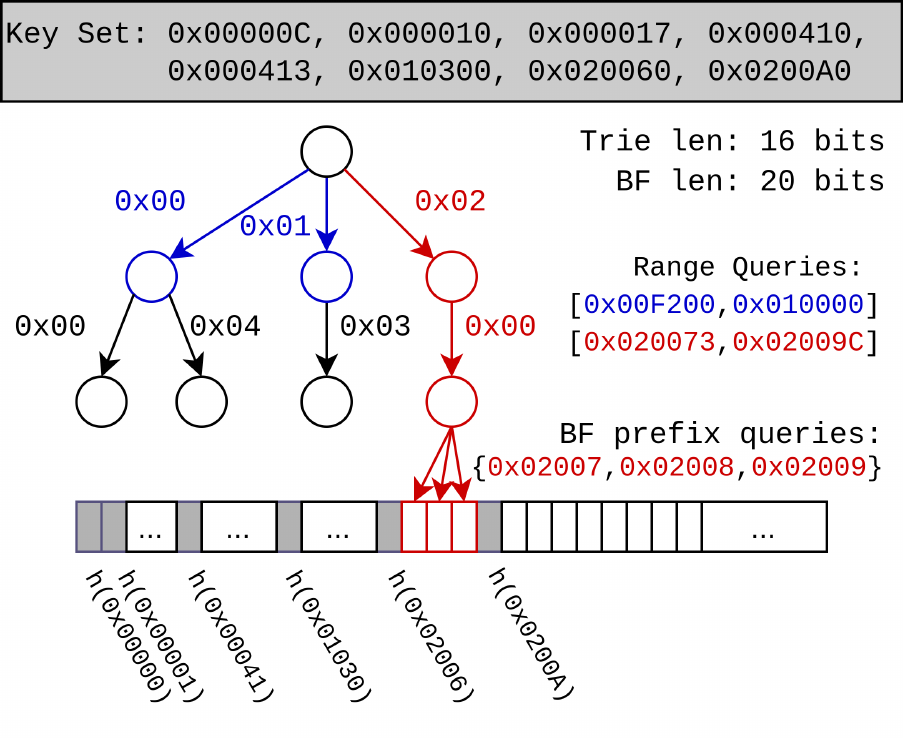}
    \vspace{-2.8em}
    \caption{An example of Proteus using 24 bit keys with a trie depth of 16 bits and a Bloom filter prefix length of 20 bits.  The blue and red show the logical paths of two empty range queries, the first of which is resolved in the trie while the second is resolved in the Bloom filter and could result in a false positive.}
    \label{fig:hybrid}
    \vspace{-0.2em}
\end{figure}

Unlike the other filters we have discussed, Proteus uses both probabilistic and deterministic encodings of the key space: an FST and a prefix Bloom filter.  
As opposed to SuRF, the FST in Proteus does not encode a unique prefix for every key, but rather all unique key prefixes of a fixed length, resulting in a FST with uniform depth.
Recall that SuRF prunes the branch for each key to the minimum length prefix that uniquely identifies it in the key set and can extend these prefixes with explicitly stored key bits. Similarly, in the Proteus FST, any trie branch encoding a single key is extended to the chosen trie depth by explicitly storing the additional key bits, rather than using the LOUDS-DS trie encoding.  A toy example of a Proteus encoding a 24 bit key space is shown in Figure \ref{fig:hybrid} using a trie depth and Bloom filter prefix length of 16 and 20 bits respectively. 

A uniform trie depth is used in part because it simplifies the modeling process, but also because we believe it to be a better use of memory.  
The intuition here is that sparse regions of the key space that use a coarse encoding are highly susceptible to false positives since they encode very large empty sub-regions.  
Similarly, encoding a densely populated portion of the key space at a coarse granularity will only result in small empty sub-regions that are less likely to cause false positives. 
SuRF requires a unique prefix for each key in order to support queries other than range emptiness such as range counts and sums, but this also imposes a minimal memory requirement and limits its usefulness for situations with strict memory constraints.  
Alternatively, a uniform depth can be adjusted to fit within any memory constraint while any leftover memory can still be put to use in the Bloom filter.  
While Proteus does not support range queries other than emptiness queries, replacing the Bloom filter with a counting Bloom filter would provide this functionality \cite{better_counting_bf}.  
The Bloom filter is assigned a prefix length between the maximum key length and the depth of the trie.  

Proteus reaps the benefits of both encodings while also mitigating their shortcomings.  
The trie is able to efficiently rule out large ranges in constant time, while the prefix Bloom filter can be positioned to catch most of the queries that would fall within the empty sub-regions encoded by the trie.
Careful choice of the Bloom filter's prefix length will also improve the prefix Bloom filter's performance independent of the trie.  
When exploring the configuration space, Proteus is free to dedicate its entire memory budget to either encoding and so can be either entirely probabilistic or deterministic depending on the context. 
Configuring Proteus amounts to choosing the prefix lengths for the trie and Bloom filter. 
The possible prefix lengths of the trie are limited by the memory budget while any memory not used in the trie can be assigned to any single prefix length using the Bloom filter.  
\begin{algorithm}[!t]
    \caption{\small The modeling process for prefix length selection.}\label{alg:model}
    \footnotesize
    \begin{flushleft}
    \hspace*{\algorithmicindent} \textbf{Input} $K$ - the key set\\
    \hspace*{\algorithmicindent} \textbf{Input} $k$ - the maximum key length in bits\\
    \hspace*{\algorithmicindent} \textbf{Input} $m$ - the memory budget in bits\\
    \hspace*{\algorithmicindent} \textbf{Input} $S$ - the set of empty sample queries\\
    \hspace*{\algorithmicindent} \textbf{Output} $(l_1, l_2)$ - prefix lengths for the best FPR \\
    \hspace*{\algorithmicindent} \textbf{Function} \texttt{trieMem($l$)} - returns size of trie with depth $l$. \\
    \hspace*{\algorithmicindent} \textbf{Function} \texttt{BFfpr($m,n$)} - returns Bloom filter FPR for $m$ bits and $n$ elements.
    \end{flushleft}
    \begin{algorithmic}[1]
        \Procedure{Model}{$K,k,m,S$}
            \State \texttt{unsigned int resolvedInTrie[$k$]}
            \State \texttt{map nRegionsQueried[$k$][$k$]}
            \Comment{stores pairs (\texttt{nRegions}, \texttt{count})}
  
            \For{$Q$ \textbf{in} $S$}
                \State \texttt{minLen} $\gets$ $lcp(K,Q)$
                \Comment{LCP is the min granularity to filter $Q$}
                \For{\texttt{tLen} $\gets 0$ \textbf{such that} \texttt{trieMem(tLen) $\leq$ $m$}}
                    \If{\texttt{tLen > minLen}}
                        \State \texttt{resolvedInTrie[tLen]++}
                    \EndIf
                    \For{\texttt{bLen} $\gets$ \texttt{tLen + 1} \textbf{up to} $k$}
                        \If{\texttt{tLen < minLen \&\& bLen > minLen}}
                            \State \texttt{nRegionsQueried[tLen][bLen].increment($I_2|L| + I_3|R|$)}
                        \EndIf
                    \EndFor
                \EndFor
            \EndFor

            \State \texttt{minFPR} $\gets 1$
            \For{\texttt{tLen} $\gets 0$ \textbf{such that} \texttt{trieMem(tLen)} $\leq$ $m$}
                \State \texttt{tFPR} $\gets$ 1 - (\texttt{resolvedInTrie[tLen] / |$S$|)}
                \If{\texttt{tFPR $\leq$ minFPR}}
                \Comment{Can be changed to \texttt{tFPR < minFPR}}
                    \State \texttt{minFPR $\gets$ tFPR}
                    \State \texttt{$(l_1, l_2) \gets$ (tLen, 0)}
                \EndIf
                \For{\texttt{bLen} $\gets$ \texttt{tLen + 1} \textbf{up to} $k$}
                    \State \texttt{bFPR} $\gets$ \texttt{BFfpr(m - trieMem(tLen),|$K_{\texttt{bLen}}$|)}
                    \Comment{Any BF can be used}
                    \State \texttt{FPR} $\gets 0$
                    \State \texttt{queries} $\gets$ \texttt{resolvedInTrie}
                    \For{\texttt{(nRegions, count)} \textbf{in} \texttt{nRegionsQueried[tLen, bLen]}}
                        \State \texttt{FPR} $\gets$ \texttt{FPR + nRegions * bFPR * count}
                        \State \texttt{queries} $\gets$ \texttt{queries + count}
                    \EndFor
                    \State \texttt{FPR} $\gets$ \texttt{FPR + (|$S$| - queries) / |$S$|}
                    \If{\texttt{FPR $\leq$ minFPR}}
                    \Comment{Can be changed to \texttt{FPR < minFPR}}
                        \State \texttt{minFPR} $\gets$ \texttt{FPR}
                        \State $(l_1,l_2) \gets$ \texttt{(tLen, bLen)}
                    \EndIf
                \EndFor
            \EndFor
            \State{\textbf{return }$(l_1, l_2)$}
        \EndProcedure
    \end{algorithmic}
\vspace{-2pt}
\end{algorithm}

\subsection{Operations}

Queries in Proteus are carried out by searching the combined structure for any members of $Q_{l_2}$ in depth-first order. 
If any prefix $x \in Q_{l_1} \cap K_{l_1}$ is found in the trie, then the prefixes $y \cap Q_{l_2}$ s.t. $x$ is a prefix of $y$ are queried in the Bloom filter.  
If any of these prefixes are present in the Bloom filter or return a false positive, the query ends and returns positive.  
If all of these prefixes return negative at the Bloom filter, the query continues to the next matching leaf node in the trie. 
If there are no more valid leaves in the trie and all queried Bloom filter regions have returned negative, the query returns negative.  
Queries that land sufficiently far from the key set are then ruled out in the trie as $lcp(Q, K)$ will be short, while queries that land closer to the key set must rely on the Bloom filter.  
In Figure \ref{fig:hybrid}, the query $Q=\texttt{[0x00F200, 0x010000]}$ (blue) is an example of the former.
The trie is searched for any member of $Q_{l_1}=\texttt{[0x00F2, 0x0100]}$, but none are found so no prefixes are queried at the Bloom filter and the query returns negative. 
However, for $Q=\texttt{[0x020073, 0x02009C]}$ (red), a matching prefix is found in the trie, $\texttt{0x0200}$, so the members of $Q_{l_2}$ with this prefix, $\texttt{\{0x02007, 0x02008, 0x02009\}}$, must all be queried at the Bloom filter.  
In this case we have that $lcp(Q,K) < l_2 \implies Q_{l_2}\cap K_{l_2} = \emptyset$, so the query will return negative so long as none of the prefixes queried at the Bloom filter result in false positives.

\subsection{Using the CPFPR Model}
\label{sec:cpfprmodel}

Proteus determines its configuration by calculating the expected FPR for each possible configuration and choosing the one resulting in the lowest FPR, as shown in Algorithm \ref{alg:model}.
This involves extracting and storing the necessary information from a set of empty sample queries and the key set, then using it to compute Equation \ref{eq:hybrid} for the desired memory budget.

\noindent \textbf{Bloom Filter FPR:} The false positive probability $p$ in Equation \ref{eq:hybrid} is dependent on the type of Bloom filter being used. We implemented Proteus using a standard Bloom filter \cite{bloom_filter} for simplicity and calculate $p$ according to Equation \ref{eq:bf}, where $n$ is the number of key prefixes in the filter, $m$ is the number of bits allocated to it, and $\lceil m/n \cdot \ln(2)\rceil$ hash functions are used.\footnote{A max limit of 32 hash functions is imposed since $m/n$ can be quite high for short prefix lengths resulting in very large hash function counts that are not practical when making multiple prefix queries. We use the MurmurHash3 and CLHASH hash functions for integer and string workloads respectively \cite{lemire2016faster}.}

\vspace{-.1in}

\begin{equation}
    p = \left( 1-e^{-\ln(2)}\right)^{\lceil m/n \cdot \ln(2) \rceil}
    \label{eq:bf}
\end{equation}

Note that both Equation \ref{eq:hybrid} and Algorithm \ref{alg:model} are AMQ-agnostic. 
The Bloom filters in our PRFs can be replaced with any AMQ, and we need only use the corresponding FPR formula.

\noindent \textbf{Count Key Prefixes:} As the Bloom filter FPR is dependent on the number of elements, we must count the number of unique key prefixes, $|K_l|$, for all possible prefix lengths $l$. 
This can be done by computing the successive LCPs of the sorted key set as each successive LCP indicates the minimum prefix length at which a key is uniquely represented. 
This is an $O(|K|)$ operation, assuming the keys are already sorted.  
In our example application, RocksDB \textemdash described in Section \ref{sec:rocksdb} \textemdash the keys are sorted internally for the filter.

\noindent \textbf{Calculate Trie Memory:} The number of unique key prefixes $|K_l|$ is also used to estimate the size of the trie at each depth $l$.
This estimation is based on the implementations of LOUDS-Sparse and LOUDS-Dense as described in \cite{surf}.
We also implemented a method to accurately calculate the trie size, but this dominated the combined modeling and construction cost of the filter and provided little benefit.  
As is, we overestimate the cost of the trie because we do not consider the memory saved by using explicitly stored key bits after a key has been uniquely represented in the FST.  
When the trie is short, this has little to no effect as very few keys will be uniquely represented. 
This error then grows with the depth of the trie, but any leftover memory is simply allocated to the Bloom filter. 
We also use this to approximate the ideal number of FST levels that should be encoded with LOUDS-Dense and LOUDS-Sparse respectively, rather than relying on a fixed ratio as SuRF does. 
This allows our FST to be even more memory-efficient than SuRF.

\noindent \textbf{Count Query Prefixes:} Here we obtain the relevant metrics from our sample queries.  
For each $Q \in S$, we must determine which trie depths will resolve the query as well as the number of regions required to cover $Q$ for each possible prefix length $l$, $|Q_l|$.  
The first of these requires $lcp(K, Q)$ which entails searching $K$ for the nearest member to $Q$. 
In the worst case, computing this for all queries is $O(|S|\log_2|K|)$, but we sort the left query bounds, which costs $O(|S|\log_2|S|)$, and start each search from the key found for the previous query. 
For the second, we simply shift the left and right bounds of $Q$ to each prefix length and take their difference.  
This is a constant amount of work for each query.

\noindent \textbf{Calculate Configuration FPRs:} Once we have gathered the above information from $K$ and $S$, we have everything we need to compute Equation \ref{eq:hybrid} for each $Q$ and configuration.  
Averaging these across a given configuration gives us our corresponding expected FPR. 
The advantage of gathering all the information first is that the false positive probabilities for the individual queries can be batched together. 
Specifically, all queries with the same $|Q_l|$ for a given prefix length $l$  have the same false positive probability and can thus be calculated together.  
The number of such calculations is then dependent on the number of queries with distinct prefix counts for each configuration.  
If range query sizes vary significantly, then most queries will have a distinct number of prefixes and the number of calculations per configuration approaches $O(|S|)$, which is not ideal.  

To counteract this, we bin the query prefix counts into $k$ bins of exponentially increasing size, where $k$ is the maximum key length in bits.
Bin $i$ contains the number of queries with prefix counts in $[2^{i-1}, 2^i)$ as well as the average number of prefix counts for those queries and bin 0 contains the number of queries resolved in the trie.
A single batch FPR calculation is then performed for each non-empty bin using the average prefix query count.
Calculating the total FPR for a given configuration therefore requires at most $k$ batch calculations. 
This significantly reduces the amount of modeling work and has little effect on the accuracy.  
This is because the probability of returning negative for an empty query decays exponentially with the number of prefix queries required.  
Despite containing more disparate values, the bins with larger ranges will still batch together queries with similar false positive probabilities.  

\begin{table}
\begin{tabular}{lcccccl}
\toprule[1pt]
$N\delta^2$ & 1 & 2 & 3 & 4 & 5 \\
\midrule
Bound & 0.00425 & 0.00132 & 0.00005 & 0.000002 & 0.0000001 \\
\bottomrule[1pt]
\vspace{8pt}
\end{tabular}
\caption{Bounds for $e^{-N\delta^2/(2p)}+e^{-N\delta^2/(3p)}$ for different values of $N\delta^2$, $p \leq 0.1$.}
\label{table:SampleSize}
\vspace{-2.5em}
\end{table}

\noindent \textbf{Sample Size:} We based our sample size on confidence intervals derived using a Chernoff bound.  
Using $N$ queries, we obtain an estimate $\hat{p}$ of the FPR of a given configuration by dividing the number of false positives found by $N$.  Assuming the $N$ queries are independently false positives with probability $p$, a standard Chernoff bound (see, e.g., Chapter 4 of \cite{mitzenmacher2017probability}) yields the following bound
for the probability that our estimate, $\hat{p}$, is within $\delta$ of $p$:
$$Pr(p \in [\hat{p} - \delta, \hat{p} + \delta]) \geq 1 - 
\min{(2e^{-2N\delta^2},e^{-N\delta^2/(2p)}+e^{-N\delta^2/(3p)})}.$$

In our setting, typically $p \leq 0.1$. 
Table \ref{table:SampleSize} provides (upper bounds for) the largest value of $e^{-N\delta^2/(2p)}+e^{-N\delta^2/(3p)})$ for $p \leq 0.1$,
for different values of $N\delta^2$. For example, with sample sizes of 10,000 and 50,000 queries, the probability that $\hat{p}$ differs from $p$ by more than $0.01$ will be at most $0.00425$ and $0.0000001$ respectively. 
Note that this is an upper bound and so the actual probability will likely be smaller in practice. Furthermore, accurately estimating the FPR of each configuration is less consequential than finding a good configuration.
So long as our estimates are close, we will end up with a configuration that is close to ideal.
We show in Section \ref{sec:modeleval:validation} the accuracy of our FPR estimates over the space of possible PRF configurations for a sample size of 10K queries.
In Section \ref{sec:modeleval:evaluation}, we compare Proteus configured using 20K sample queries against the state-of-the-art on diverse workloads.

\begin{table}
\begin{tabular}{lcccccl}
\toprule[1pt]
& \multirow{3}{*}{\parbox{0.12\linewidth}{\centering \textit{Count Key Prefixes}}} & \multirow{3}{*}{\parbox{0.07\linewidth}{\centering \textit{Calc. Trie Mem.}}} & \multirow{3}{*}{\parbox{0.12\linewidth}{\centering \textit{Count Query Prefixes}}} & \multirow{3}{*}{\parbox{0.1\linewidth}{\centering \textit{Calc. Config FPRs}}} & \multirow{3}{*}{\parbox{0.09\linewidth}{\centering \textit{Build Filter}}} & \multirow{3}{*}{\parbox{0.07\linewidth}{\centering \textit{Total}}} \\
& & & & & & \\
& & & & & & \\
\midrule
\textbf{1PBF}          & 32 & -      & 9     & 1      & 3139      & 3181 \\
\textbf{2PBF}          & 32 & -      & 81    & 884   & 3239       & 4236 \\
\textbf{Proteus}          & 32 & 1     & 71    & 1      & 3130      & 3235 \\
\textbf{SuRF}           & \multicolumn{4}{c}{-}      & 482       & 482 \\
\textbf{Rosetta}        & \multicolumn{4}{c}{3}      & 4775      & 4778 \\
\bottomrule[1pt]
\vspace{5pt}
\end{tabular}
\caption{Breakdown and comparison of filter construction times, including modeling. Values are rounded up to nearest millisecond.}
\label{table:FilterConstructionCostBreakdown}
\vspace{-20pt}
\end{table}

\noindent \textbf{Modeling Cost Breakdown:} Table \ref{table:FilterConstructionCostBreakdown} shows a breakdown of these costs for 10M normally distributed keys, a sample of 20K correlated empty queries, and a memory budget of 10 BPK.
This workload is designed to maximize the number of possible configurations for Proteus and consequently the computation required for modeling.  
We use normally distributed keys to increase the number of viable configurations to test as there will be more common key prefixes, thereby making the trie more compact.
The queries are correlated to the keys just enough that most will not be resolved in the trie which increases the number of calculations required to compute the expected FPR for each configuration.  
Lastly, we use range sizes uniformly distributed between $2$ and $2^{20}$ to have a large number of distinct prefix counts. 
2PBF uses a maximum range size of $2^{15}$ due to values overflowing when computing the binomial coefficient in Equation \ref{eq:dpbf} for queries with a large number of prefixes.

\begin{figure*}[h]
    \centering
    \begin{subfigure}[b]{0.28\textwidth} 
        \includegraphics[width=\linewidth]{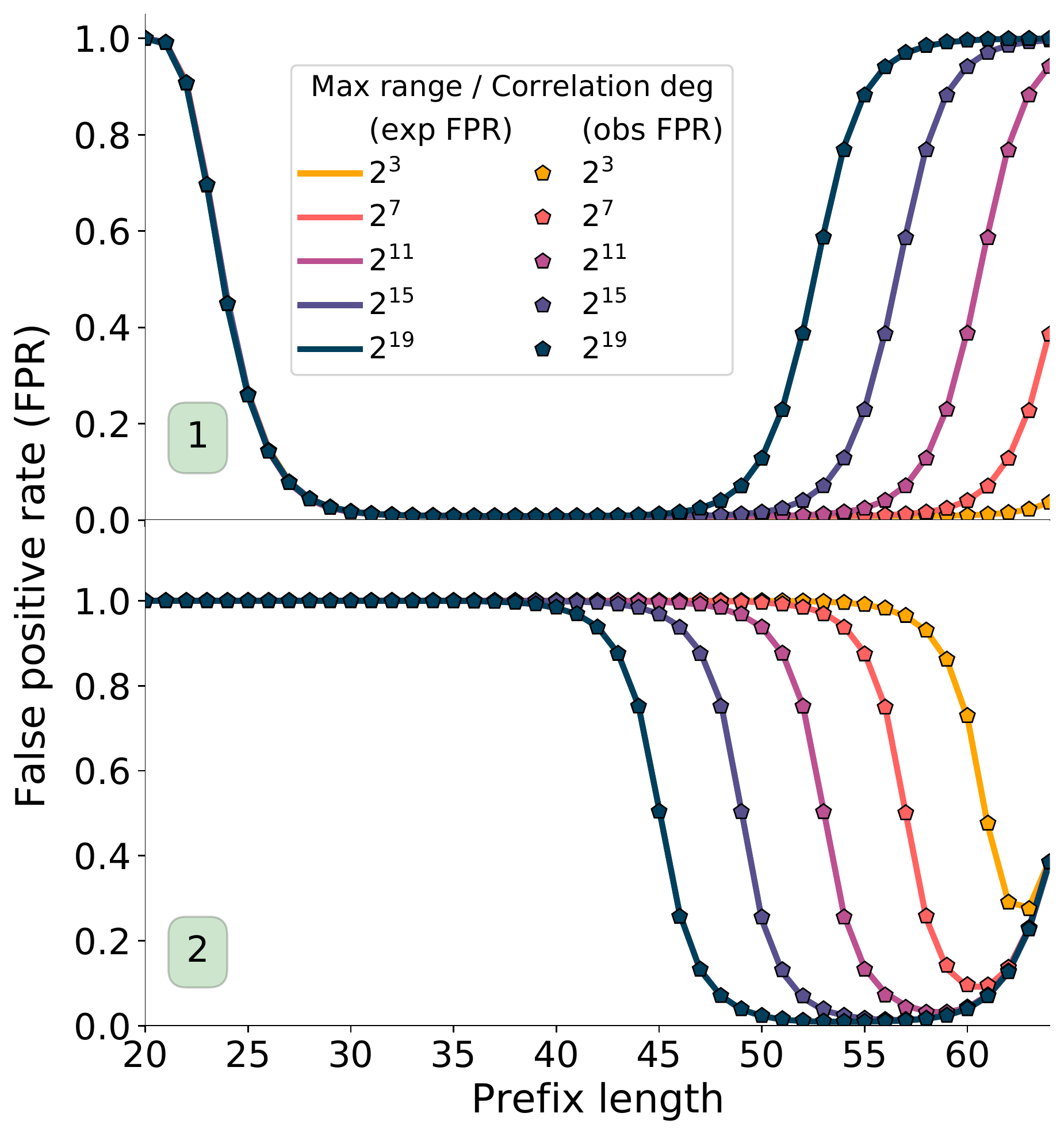}
        \caption{1PBF}
        \label{fig:pbf_validation}
    \end{subfigure}
    \hfill
    \begin{subfigure}[b]{0.3\textwidth}
          \includegraphics[trim=0 0 430 0, clip=true, width=\linewidth]{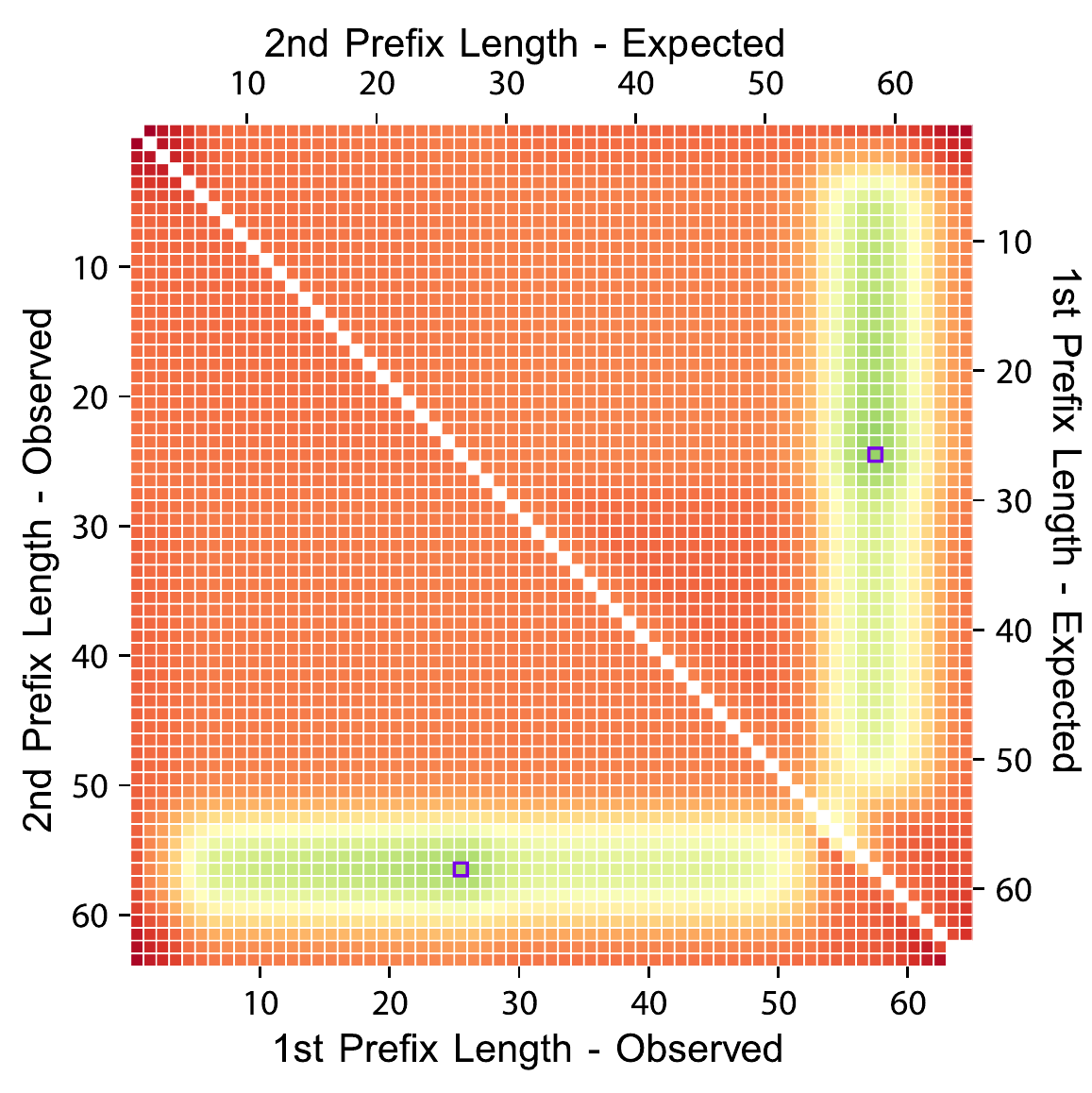}
        \caption{2PBF}
        \label{fig:heat_map_dpbf}
    \end{subfigure}
    \hfill
    \begin{subfigure}[b]{0.37\textwidth}
      \includegraphics[trim=350 0 0 0, clip=true, width=\linewidth]{Figures/Figure4bc.pdf}
        \caption{Proteus}
        \label{fig:heat_map_hypr}
    \end{subfigure}
    \caption{The CPFPR model accurately predicts the FPR for all possible designs of different Protean Range Filters.}
    \vspace{3pt}
\label{fig:meval}
\end{figure*}

Looking at the results in Table \ref{table:FilterConstructionCostBreakdown}, we see that the modeling time for 1PBF (\textasciitilde42ms) is about two orders of magnitude smaller than its construction time (\textasciitilde3s), which is just that of a standard Bloom filter.
Proteus's modeling is a modestly more expensive (\textasciitilde100ms) but is still dominated by the construction time (\textasciitilde3s). 
It is worth noting that, without the binning, calculating the expected FPRs for each configuration becomes the dominant factor for all of our PRFs, in the worst case. 
With the binning, the combined modeling and build cost of both 1PBF and Proteus is comparable to the construction cost of a standard Bloom filter.
However, the modeling cost alone for 2PBF is comparable to the construction cost for a Bloom filter, even with reduced range size.
This is for a number of reasons. 
While Proteus's potential configurations are limited by the cost of its trie, 2PBF's choice of $l_1$ has no such limit.
2PBF therefore considers all combinations of $l_1 < l_2 \in [1,64]$ for multiple memory allocations.
An exhaustive search of all possible memory allocations is infeasible, so we implement 2PBF to test 2 asymmetric allocations (60-40/40-60) and a symmetric allocation. 
Furthermore, the probabilistic nature of the first filter results in many possible outcomes for each query, all of which must be considered when calculating the query's false positive probability. 
This puts 2PBF's modeling time (\textasciitilde1s) on the same order of magnitude as it's construction time (\textasciitilde3s). 
Note that this is a worst case workload for modeling time but not filter build time. 
This is because 2PBF only ends up using a single Bloom filter while Proteus does not use a trie.
However, the worst case build time for Proteus will not be significantly larger since the FST can be built very quickly, as shown by SuRF's build time.
2PBF's build time can as much as double if it has to build a second Bloom filter.

\section{Model Validation}
\label{sec:standalone}

\begin{figure*}
    \centering
    \includegraphics[width=\textwidth]{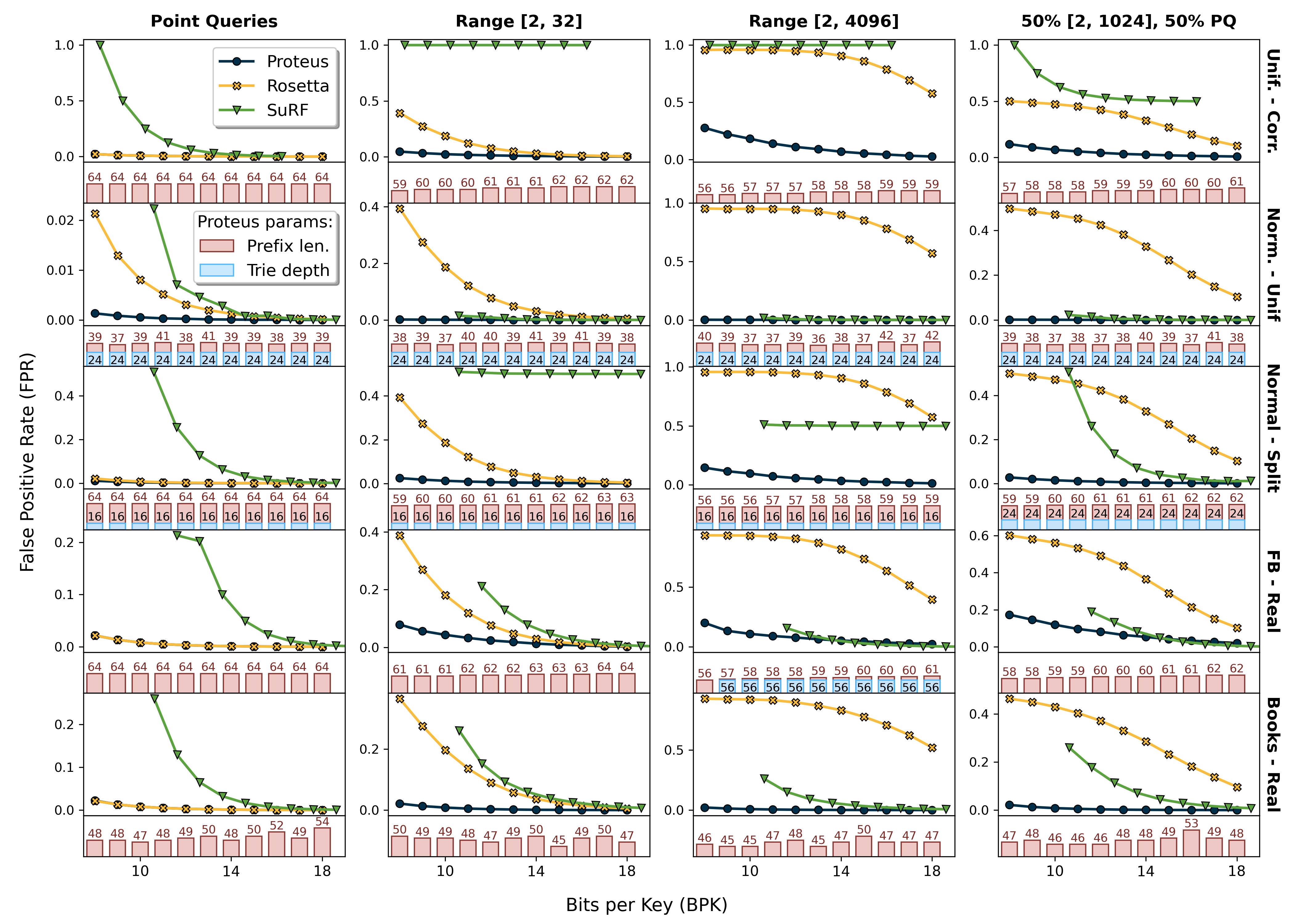}
    \vspace{-2em}
    \caption{Proteus optimally configures its design on diverse workloads with varying range sizes and memory budgets.}
    \label{fig:standalone}
    \vspace{3pt}
\end{figure*}

Here we demonstrate the accuracy of the described CPFPR models and evaluate our PRFs (1PBF, 2PBF, and Proteus) as in-memory standalone filters. 
We show that Proteus selects optimal designs across a variety of workloads and achieves better FPRs than state-of-the-art AREs due to its broader design space.
In Section \ref{sec:rocksdb}, we evaluate these AREs in RocksDB with end-to-end system metrics.   
We use 64-bit unsigned integers for our experiments in Section \ref{sec:standalone}
and Section \ref{sec:rocksdb} and focus on strings in Section \ref{sec:strings}. 

\noindent \textbf{Datasets:} The real world datasets come from the Searched on Sorted Data (SOSD) benchmark for index structures \cite{sosd-vldb, sosd-neurips}.

\begin{itemize}[leftmargin=*]
    \item \textsc{Uniform:} Keys are generated uniformly from \texttt{[0, $2^{64}$ - 1]}.
    \item \textsc{Normal:} Keys are generated with a mean of $2^{63}$ and a standard deviation of $2^{64}\cdot0.01$.  
    \item \textsc{Books:} Amazon booksale popularity data for 800M books. This data has a fairly heavy skew with many more low popularity scores than high.   
    \item \textsc{Facebook:} A set of 200M upsampled Facebook user IDs.  This data is fairly dense and covers a relatively small range of values with uniformly distributed gaps.  
\end{itemize}

\noindent \textbf{Workloads:} We test variations of YCSB Workload E, a range-scan intensive workload. Queries are of the form \texttt{[left, left+offset]}, where \texttt{offset} is chosen uniformly at random from \texttt{[2, RMAX]} unless otherwise stated. For point queries, \texttt{offset} is set to 0.
\begin{itemize}[leftmargin=*]
    \item \textsc{Uniform:} \texttt{left} is taken uniformly at random from \texttt{[0, $2^{64}-\texttt{RMAX}$]}.
    \item \textsc{Correlated:} A \texttt{key} is chosen uniformly at random from the dataset and then \texttt{left} is chosen uniformly at random from \texttt{[key+1, key+CORRDEGREE]}. We use a default \texttt{CORRDEGREE} of $2^{10}$.
    \item \textsc{Split:} An even split of \textsc{Uniform} and \textsc{Correlated} queries, similar to the particle physics workload mentioned in Section \ref{sec:intro}.
    \item \textsc{Real:} For a real world dataset, we uniformly sample 10M integers to use as keys and another 1M integers to use as the \texttt{left} bounds. 
\end{itemize}

\noindent \textbf{Experimental Setup:} All experiments were run on Linux kernel 5.13.12-arch1-1 with an AMD Ryzen 7 1800X 8-core processor, 16GB RAM, and 1TB Samsung 850 EVO SSD. For each experiment, 1M queries were executed serially on a single thread with a sample of 20K queries and a data set of 10M keys.

\subsection{Model Accuracy}
\label{sec:modeleval:validation}

We validate the accuracy of the CPFPR models for 1PBF, 2PBF, and Proteus by comparing the expected FPR \textemdash as calculated by the corresponding CPFPR model \textemdash with the observed FPR on all possible designs in the respective filter design spaces.
We use a sample size of 10K queries for these experiments, demonstrating our accuracy at the lowest $N\delta^2$ in Table  \ref{table:SampleSize}. 

\noindent \textbf{1PBF:} 
We run two experiments to highlight the impact of each of our contextual parameters, range size and correlation between keys and queries, as shown in Figure \ref{fig:pbf_validation}.
The top graph varies \texttt{RMAX} on \textsc{Uniform-Uniform}, while the bottom graph varies \texttt{CORRDEGREE} on \textsc{Uniform-Correlated}.
The \texttt{RMAX} for the bottom graph is fixed at $2^7$ and the prefix length on the x-axis represents the different possible designs for 1PBF.

In Figure \ref{fig:pbf_validation}.1, we see that the observed FPR quickly increases once the prefix length passes $64 - \log_2 \texttt{RMAX}$. 
Before this threshold, a given range query will not require more than 2 regions to be queried in the prefix Bloom filter as the range queries are all smaller than the regions encoded the prefix filter.  
Since we use \textsc{Uniform} keys and queries, the significant majority of queries do not fall close to keys. 
As such, empty sub-regions only become an issue for prefix lengths shorter than 30.  
The same effect becomes more prevalent in Figure \ref{fig:pbf_validation}.2 where we see that any prefix length shorter than $64 - \log_2 \texttt{CORRDEGREE}$ is affected by empty sub-regions. 
Since Figure \ref{fig:pbf_validation}.b uses a range size of $2^7$, any prefix length longer than 57 will also result in more false positives due to range size.  
When $\log_2 \texttt{CORRDEGREE} > 7$, the prefix filter must contend with false positives resulting from both empty sub-regions and range size.  
In both experiments, the 1PBF CPFPR model is able to accurately capture the effects of both conditions as they pertain to the FPR.

\noindent \textbf{2PBF \& Proteus:} 
For both of our PRFs that use two prefix lengths, we focus on a situation that calls for multiple prefix lengths, as described in Section \ref{sec:modeling}. 
We use \textsc{Normal-Split} with short range \textsc{Correlated} and long range \textsc{Uniform} queries to necessitate the use of two prefix lengths.  
The FPRs for each design are shown in Figure \ref{fig:heat_map_dpbf} and \ref{fig:heat_map_hypr} for 2PBF and Proteus respectively.
The expected FPRs are shown in the upper right triangle of the matrix while the corresponding observed FPRs are reflected in the lower left. 

For both filters, we correctly predict the optimal design, and accurately predict the range of FPRs over the entire design space.
The optimal design for 2PBF is uses prefix lengths of 26 and 57 bits while Proteus's optimal design uses a 24 bit trie and a prefix length of 58 bits.
The corresponding expected and observed FPRs are 11.4\% and 12.3\% for 2PBF and 5.17\% and 4.91\% for Proteus respectively.  
These values may seem high in terms of the FPRs typically achieved by AMQs, but this is also a highly adversarial case which current AREs are unable to handle.  
We re-visit this use case in Section \ref{sec:rocksdb:etoe} in the context of end-to-end system performance.   
The gray region corresponds to the part of the parameter space where the trie is too large for the memory budget (10 BPK).
Despite its more limited design space, Proteus achieves lower FPRs than 2PBF.

Figure \ref{fig:heat_map_dpbf} shows the results for the best memory division between the two Bloom filters, a 50-50 split. 
This works best as our \textsc{Split} query distribution is evenly split between small, key-correlated queries and large, uniformly distributed queries.  
2PBF also considers asymmetric memory allocations, as described in Section \ref{sec:PRFs}.

Both Proteus and 2PBF fully encompass 1PBF's design space and will always achieve an equivalent or lower FPR.  
Moreover, even though Proteus and 2PBF occupy slightly different design spaces, in the situations where a second Bloom filter is helpful, it is outperformed by Proteus's trie.

\vspace{-1em}
\subsection{Optimizing Across the Design Space}
\label{sec:modeleval:evaluation}

We now demonstrate Proteus's ability to select optimal designs across a variety of workloads.  
We also contrast Proteus flexibility against state-of-the-art AREs Rosetta and SuRF for each, as shown in Figure \ref{fig:standalone}.
Each row in Figure \ref{fig:standalone} shows the results for a \textsc{Dataset-Workload} pair for point queries, small range queries, large range queries and a combination of point and range queries.
The SuRF results show the lowest FPR for all possible configurations of real and hash-suffix bits, but in practice users will need to implement a policy to choose the appropriate SuRF configuration.
In these experiments, Proteus and Rosetta both use 20K empty sample queries which gives us a bound of $0.00425$ for $\delta = 0.1$, as per Table \ref{table:SampleSize}. 
This will give us higher confidence in the optimality of Proteus's chosen design as we compare its performance across workloads.

\noindent \textbf{Effective Navigation of Design Space}: For nearly all use cases, Proteus is able to choose a design that achieves a low FPR.
This is less true for large \textsc{Correlated} queries \textemdash an adversarial case for any prefix-based filter \textemdash as Proteus must rely entirely on a Bloom filter design.
Even so, Proteus is able to achieve a much lower FPR than any state-of-the-art ARE by picking a prefix length that balances the number of prefixes per query and the number of range queries distinguishable from the key set.
In situations where SuRF and Rosetta are optimal, Proteus takes on a similar design to the respective filter and achieves similar performance.
For instance, Rosetta and Proteus are virtually indistinguishable in terms of FPR in point query workloads.
Similarly, Proteus and SuRF achieve very similar FPRs on \textsc{Facebook-Real} as the keys lie in a narrow range which causes both Proteus and SuRF to have extremely deep tries.

\noindent \textbf{Impact of Restricted Design Spaces}: However, if the optimal design lies outside the restricted design spaces of SuRF or Rosetta, then the corresponding filter's performance will be limited.
For example, consider the \textsc{Uniform-Correlated} small range query workload. 
Despite only a small increase in range size from the point query workload, a full length prefix filter is no longer optimal for lower memory budgets as the benefit from distinguishing all queries is outweighed by the benefit of querying fewer prefixes, as corroborated by Proteus's design.
Even so, Rosetta will always dedicate the majority of its memory to the full length prefix filter.
For all point query workloads as well as the mixed query workloads for \textsc{Uniform-Correlated} and \textsc{Normal-Split}, SuRF achieves its best FPR with the use of hash-suffix bits and only achieves good FPR with high memory budgets.
In such cases, a Bloom filter is a more efficient use of memory and can be tuned to optimize for arbitrary range sizes, in contrast to the hash-suffix bits which are only used for point queries.
We can also observe SuRF's minimum memory requirement across the various workloads. In most cases, it requires 11-12 BPK, while Proteus can always achieve equivalent if not better performance at 8BPK.

\noindent \textbf{Additional Benefits of the Hybrid Design Space}: The combination of complementary design elements in Proteus allows it to achieve better FPR than designs which rely only on a single technique.
Even for some point query workloads, such as \textsc{Normal-Uniform}, a hybrid design can leverage a short, memory-efficient trie to achieve a better FPR-memory tradeoff than a standard Bloom filter.
Furthermore, on \textsc{Split} workloads, Proteus is able to gracefully handle both types of queries, but more brittle structures may have to sacrifice performance on a certain portion of the queries.
For instance, on the mixed \textsc{Normal-Split} workload at a low memory budget, Rosetta and SuRF can only filter the \textsc{Correlated} point queries and the \textsc{Uniform} range queries respectively.

\vspace{-1em}
\section{System Evaluation: RocksDB}
\label{sec:rocksdb}

We integrate Proteus into RocksDB \textemdash a popular key-value store \textemdash and demonstrate that it improves end-to-end range query performance by up to 3.9x and 3.3x over Rosetta and SuRF respectively across a variety of workloads, with consistently better performance at low BPKs (8-12).  
This speed-up is due to a reduction in I/O operations as a result of the lower FPR achieved.
Furthermore, we show that the additional cost of modeling in filter construction does not significantly impact the end-to-end performance of RocksDB and that Proteus is able to adapt smoothly to changes in the query workload distribution, unlike Rosetta and SuRF which suffer drastic declines in performance on certain adversarial distributions.

\subsection{Proteus System Integration}

RocksDB uses an LSM tree architecture which organizes data on disk into levels of increasing size, where each level $L_i$ (except $L_0$) is range partitioned into multiple sorted runs or Static Sorted Table (SST) files that occupy disjoint key ranges.
The SST files in $L_0$ are directly flushed to disk from MemTables \textemdash in-memory structures that buffer writes \textemdash and thus typically have overlapping key ranges.
Static filters (e.g. Bloom filters) are built on every SST file to reduce unnecessary accesses for non-existent keys.
When a level $L_i$ reaches its maximum capacity, RocksDB selectively merges SST files from $L_i$ into $L_{i+1}$, triggering the construction of new filters on the merged data in the new $L_{i+1}$ SST files.
This process is called compaction.

\noindent \textbf{Range Query Implementation:} Similar to \cite{rosetta}, we modify the RocksDB closed \texttt{Seek} logic to first check all filters for the existence of keys in the queried range. 
If all filters return false, then \texttt{Seek} returns an invalid iterator.
For the filters that return true, RocksDB proceeds to retrieve the smallest keys from the associated SST files that are greater than or equal to the lower query bound.
This is done by binary searching over the index block which stores min/max information in each SST file and fetching the corresponding data block.
If the global smallest key is smaller than the upper query bound, an iterator pointing to that key is returned. 
Otherwise, an invalid iterator is returned for the empty range query.

\noindent \textbf{Sample Query Queue:} Since Proteus (and Rosetta) need sample queries, we create a fixed size query queue and seed it with an initial query sample.
Older queries are evicted with a FIFO policy.
This changing set of sample queries is used in conjunction with the keys in each SST file to determine the optimal filter design for each SST file at construction time.
In our experiments, we use a queue size of 20K queries (\textasciitilde320KB) and update the queue with every 100th executed empty query.

\begin{figure*}[ht]
    \centering
    \includegraphics[trim=0 15 0 0 , clip=true, width=\linewidth]{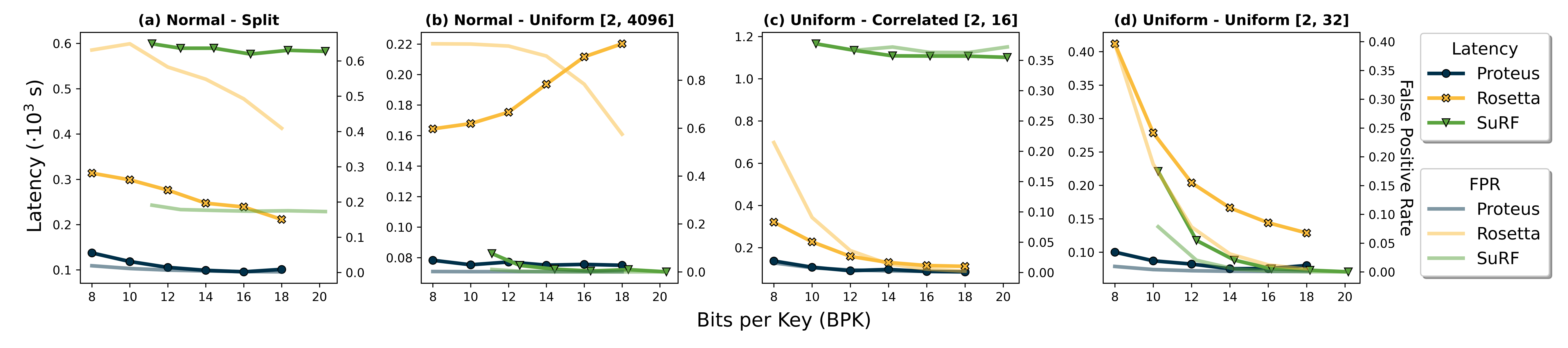}
    \vspace{-1.2em}
    \caption{Proteus improves end-to-end RocksDB performance on low memory budgets across diverse workloads.}
    \label{fig:rocksdb_etoe}
\end{figure*}

\noindent \textbf{Tuning RocksDB:} Since all filters have to be queried during a closed \texttt{Seek}, we curtail this CPU cost by tuning RocksDB to fit more data in each SST file, thereby reducing the number of filters queried. 
This also helps to control the overall cost of modeling from filter construction when there are heavy compactions. 
We increase the SST file size from the default of 64MB to 256MB and scale up the size of $L_1$ and the MemTables to maintain the same number of SST files that fit in them with default settings.\footnote{\texttt{write\_buffer\_size = 256MB, max\_bytes\_for\_level\_base = 1024MB}}
Similarly, we selectively enable data compression for certain levels.
We leave the few SST files in $L_0$ and $L_1$ uncompressed to maintain the speed of MemTable flushes and $L_0 \rightarrow L_1$ compactions.
For SST files in $L_2$, we use LZ4 compression \textemdash a light-weight compression algorithm recommended by RocksDB \cite{rocksdb_compression} that balances CPU cost and compression size.
We use heavy-weight ZSTD compression for SST files in $L_3$ onwards as they are less frequently modified and contain bulk of the data in the LSM tree.
We ensure that the filters are cached in the RocksDB block cache\footnote{\texttt{cache\_index\_and\_filter\_blocks = true,\\ pin\_l0\_filter\_and\_index\_blocks\_in\_cache = true}} and assign RocksDB 6 background threads for flushes and compactions.

\subsection{Experimental Setup}
We use the same datasets, range query workloads, and machine described in Section \ref{sec:standalone} for experiments carried out in RocksDB v6.20.3.
For each 8 byte integer key, we generate an associated 512 byte value.
The first half of all values are zeroed out, while the second half is randomly generated which yields a constant compression ratio of 0.5.
To ensure that all experiments start from a consistent LSM tree state, we manually flush the MemTable after populating the initial, empty database, and wait for all background compactions to finish before executing the benchmark.
In Section \ref{sec:rocksdb:etoe}, the initial database has 50M key-value pairs which yields a 4 level (\textasciitilde14GB compressed) LSM tree with \textasciitilde70 SST files.
In Section \ref{sec:rocksdb:robustness}, we first insert 20M key-value pairs to get a 3 level (\textasciitilde6GB compressed) LSM tree with \textasciitilde40 SST files, and subsequently \texttt{Put} an additional 40M key-value pairs over the course of the experiments.
In both cases, $L_0$ is empty in the initial database state as we set RocksDB to compact all $L_0$ SST files to $L_1$ for sake of consistency.
Lastly, we warm the RocksDB block cache (1GB) and the OS page cache by running 1M uniformly distributed point queries of existing keys to ensure that all indexes and filters are loaded into memory.

\subsection{End-to-End Performance}
\label{sec:rocksdb:etoe}

\begin{figure}[t!]
    \centering
    \includegraphics[width=0.48\textwidth]{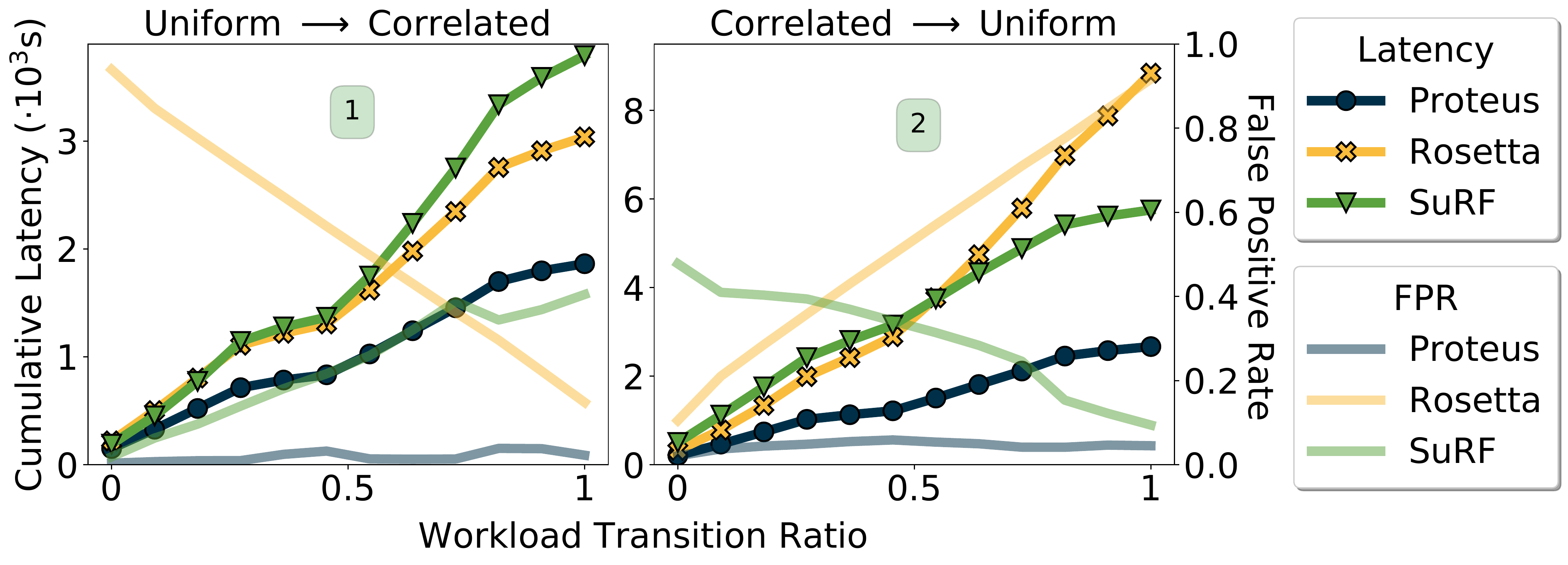}
    \vspace{-2em}
    \caption{Proteus is robust against extreme workload shifts.}
    \label{fig:robustnessRtoS_validation}
\end{figure}

\begin{figure}[t]
    \centering
    \includegraphics[width=0.48\textwidth]{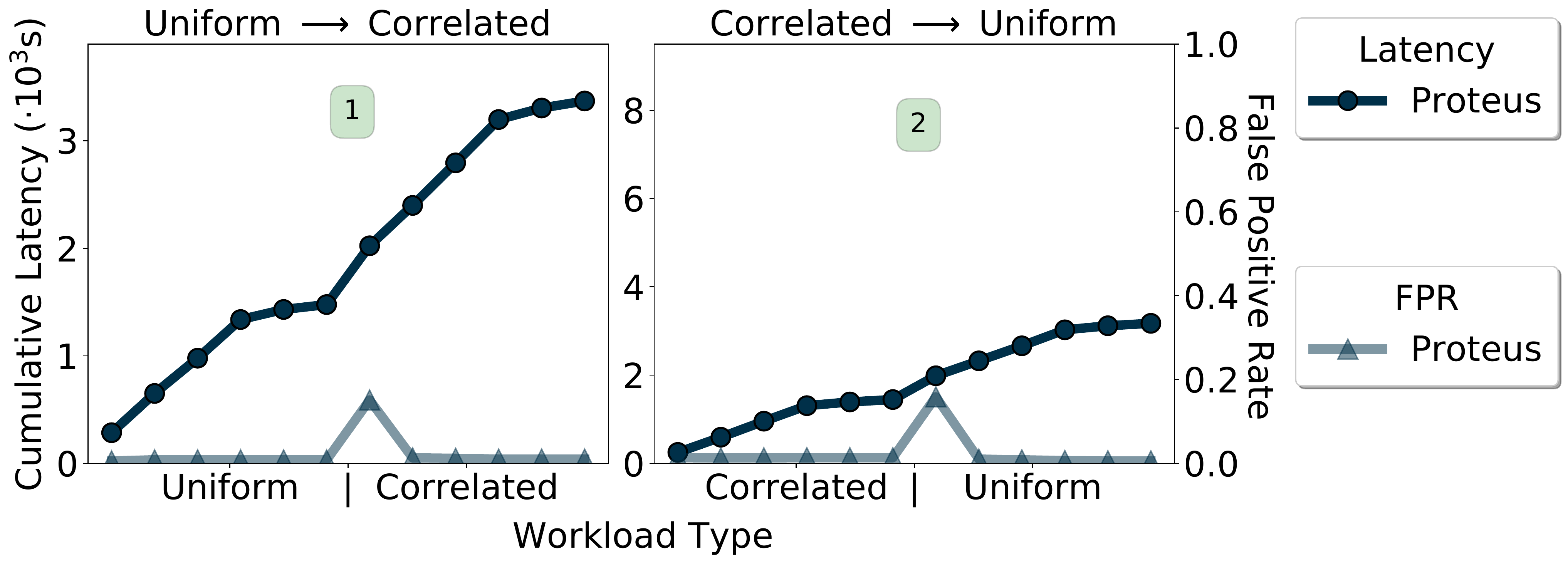}
    \vspace{-2em}
    \caption{Proteus is robust to immediate, extreme workload shifts.}
    \label{fig:extreme_robustnessRtoS}
\end{figure}

We measure the end-to-end range query performance in terms of workload execution latency for Proteus, SuRF, and Rosetta on four use cases targeting different points in the design space. 
As shown in Figure \ref{fig:rocksdb_etoe}, Proteus achieves the lowest latency across all workloads on low memory budgets, improving upon SuRF and Rosetta by as much as 3.3x and 3.9x respectively.
In RocksDB, one of the primary sources of latency is I/O when accessing the SST files.
As such, a lower FPR generally results in lower latency because we can avoid unnecessary I/O. 
However, situations such as thrashing in RocksDB's internal cache or excessive prefix queries during Bloom filter probes can result in higher latency despite a low FPR.  
For example, Rosetta improves its FPR on \textsc{Normal-Uniform} as its memory budget increases by using more hash functions per Bloom filter, but the resulting latency increases.
This is because a large range query requires Rosetta to query many prefixes and the CPU cost for each is proportional to the number of hash functions used. 
Proteus is able to frequently avoid this as the trie limits - and sometimes eliminates all - prefix queries made at the Bloom filter.
SuRF has an effectively constant computational cost for queries, but can suffer due to pressure on RockDB's internal cache.  
This can be seen in both the \textsc{Normal-Split} and \textsc{Uniform-Correlated} workloads.  
We observed that SuRF puts more pressure on RocksDB's internal cache which results in thrashing after passing a certain FPR threshold (\textasciitilde0.1 to 0.2 in our experiments).\footnote{\texttt{rocksdb.block.cache.add} shows that more data blocks are added to the cache when using SuRF compared to Proteus with similar FPR.}
This thrashing severely impacts the overall system latency.
We also observed that SuRF's memory footprint varies by as much as 3 BPK across SST files, while Proteus and Rosetta maintain consistent BPKs and do not result in thrashing.  Naturally, I/O savings would be magnified for  larger datasets.

\subsection{Robustness to Shifting Query Distribution}
\label{sec:rocksdb:robustness}

We now evaluate the robustness of Proteus: its ability to maintain good end-to-end performance when the workload shifts.
We examine incrementally shifting workloads which mimic real-life applications with temporal skew, such as Wikipedia.
As shown in Section \ref{sec:modeling}, a range filter's performance is primarily affected by the relative positioning between the data keys and the queried keys. Thus, changing the queries or the data are essentially equivalent in terms of impact on filter performance and so we focus on changing the query distribution to be able to  control the relative key-query proximities.
We test with workloads that gradually shift between large range \textsc{Uniform} queries and small range \textsc{Correlated} queries which favor shorter and longer prefix lengths respectively.
To magnify the difference in Proteus designs, we maintain a \textsc{Normal} key distribution when shifting from long \textsc{Uniform} queries to short \textsc{Correlated} queries as the trie chosen for \textsc{Uniform} queries is ineffective for \textsc{Correlated} queries and take memory away from the Bloom filter. Similarly, we maintain a \textsc{Uniform} key distribution for a short \textsc{Correlated} to long \textsc{Uniform} query transition which precludes the use of a trie for long \textsc{Uniform} queries.

For each workload, we define the workload transition ratio as the probability of executing a query from the end query distribution.
We test 60M closed \texttt{Seeks} with a workload transition ratio increasing linearly from 0 to 1.
We start with an initial database of 20M keys and uniformly interleave 40M \texttt{Puts} with the 60M \texttt{Seeks} to trigger periodic compactions and  construction of new filters.

In Figure \ref{fig:robustnessRtoS_validation}, we show the cumulative latency as the respective workloads transition from one type of query to another, and report the FPR for every batch of 5M \texttt{Seeks}.
Proteus is resilient to the extreme workload shifts and is able to instantiate new designs to maintain a consistently low \texttt{Seek} latency.
As shown in Figure \ref{fig:robustnessRtoS_validation}.1 and \ref{fig:robustnessRtoS_validation}.2, Proteus has a smooth increase in cumulative latency which stems from the low FPR maintained as the workload shifts.
This is because Proteus can configure itself accurately by relying on the sample query queue to provide an up-to-date view of the query workload.
The end-to-end behavior observed also highlights that the additional cost of modeling for Proteus during filter construction does not impact the overall performance despite heavy ongoing compactions (\textasciitilde15-20 for each 5M \texttt{Seek} batch).
In contrast, the latencies for SuRF and Rosetta increase sharply when the workload transitions past 0.5 for \textsc{Uniform\textrightarrow Correlated} (Figure \ref{fig:robustnessRtoS_validation}.1) and \textsc{Correlated\textrightarrow Uniform} (Figure \ref{fig:robustnessRtoS_validation}.2) respectively.
Due to their restricted design spaces, Rosetta and SuRF can only effectively handle one of the two types of queries.
We observe the impact of their brittle designs in the FPR which decreases as the workload shifts.

We repeated the same experiments for Proteus with an immediate change in query distribution, simulated by not mixing the two distributions, with results shown in Figure \ref{fig:extreme_robustnessRtoS}. We observe a larger increase in FPR and latency after the drastic transition since the filter designs are not optimal for the new query distribution, but the decrease in performance is temporary as the filters are rebuilt using the updated query cache, giving robust performance.
\vspace{-0.3em}

\section{Variable Length Keys}
\label{sec:strings}

\noindent
Database workloads commonly include variable length keys, which often arise from concatenations of various metadata \cite{facebook_workloads, idreos2020key}.
In this section, we show how Proteus can be used with variable length keys and demonstrate that the CPFPR model extends to any key length.
We also show that Proteus reduces end-to-end query latency in RocksDB by as much as 5.3x vs. SuRF on a real-world string dataset.

\subsection{Extending the Model and Filter}

\textbf{Modeling:} Shifting from a fixed length integer key space to a variable length key space in the CPFPR model is equivalent to changing the total ordering from a less-than-or-equal relation to a lexicographical relation. For longer keys, Proteus also has more potential designs due to the larger range of prefix lengths. In the context of static filters, the length of the longest key in the dataset is known and therefore Proteus's design space is well defined.

\noindent
\textbf{Filter Operation:} The trie portion of Proteus handles string keys without requiring any modifications. On the other hand, variable length keys give rise to an exploding number of prefixes to query in the prefix Bloom filters. 
In addition, every Bloom filter query inherently increases the probability of a false positive. 
Proteus achieves low FPR and query time by padding short keys and queries with trailing null bytes to a chosen prefix length, thus mapping the key space onto a fixed-length key space.
This means that the prefix Bloom filter does not distinguish between short keys and their padded equivalents, which will result in false positives if the application does not make the same assumption.
Finally, we changed the Bloom filter hash function from MurmurHash3 to CLHASH which can handle strings \cite{murmurhash, lemire2016faster}.

\subsection{Validation and Evaluation}
\noindent \textbf{Experimental Setup:} We run in-memory and RocksDB benchmarks similar to Section \ref{sec:modeleval:evaluation} and Section \ref{sec:rocksdb:etoe} respectively.
For in-memory experiments, we generate three datasets of fixed-length string keys of size 80, 200, and 1440 that conform to either a \textsc{Uniform} or \textsc{Normal} distribution.
\textsc{Uniform} keys are concatenations of uniformly generated key bytes up to the specified key length, while \textsc{Normal} keys are normally distributed around the middle of the key space with standard deviation $\sigma = 0.01\cdot2^{64}$. 
Specifically, the mean key is defined to be the string with a most significant byte value of 128 followed by null bytes up to the key length. 
We also generated \textsc{Uniform}, \textsc{Correlated}, and \textsc{Split} synthetic string workloads with \texttt{RMAX} $2^{30}$ and \texttt{CORRDEGREE} $2^{29}$.
The in-memory experiments were run with 10M keys, 1M queries, and 20K sample queries.
For RocksDB experiments, we use an internet domain dataset comprising \textasciitilde31M crawled \texttt{.org} domains \cite{domains}.
The domains are 5 to 253 bytes long with a median length of 21 bytes and follow a log-normal distribution ($R^2 = 0.98$).
The initial database was populated with 20M domain keys and 512 byte random values, resulting in a 3 level (\textasciitilde6GB compressed) LSM tree with \textasciitilde30 SST files.
Another 10M random domains were used to generate queries with \texttt{RMAX} $2^{30}$, as with our other \textsc{Real} workloads.
To control the distribution of \texttt{RMAX}, we pad the dataset with null bytes to the max key length.
Note that this does not affect the performance of SuRF as it only considers the keys up to their unique prefixes which will be unchanged by the padding.
For Proteus, the padding would ideally be done on a per-SST file basis to avoid unnecessary padding and modeling of designs with longer prefix lengths.
These costs are incurred at construction time and not measured in the read-only experiment.

\begin{figure*}
    \centering
     \begin{subfigure}[h]{0.74\textwidth}
         \centering
         \includegraphics[width=\textwidth]{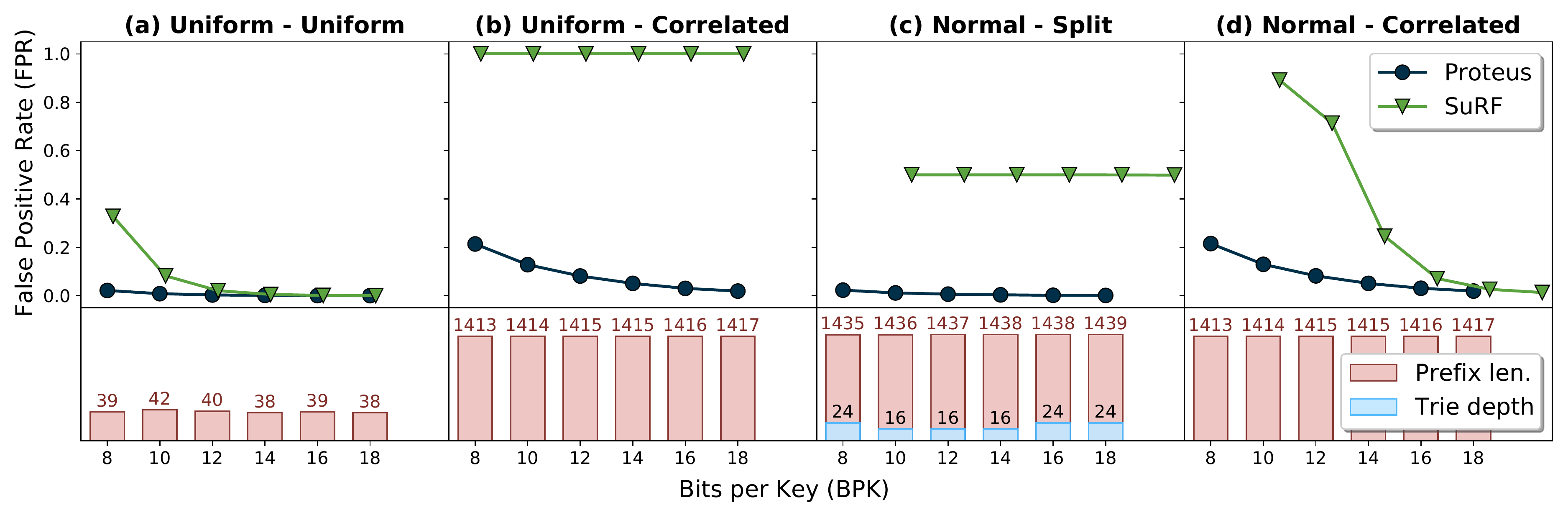}
     \end{subfigure}
     \begin{subfigure}[h]{0.24\textwidth}
         \centering
         \includegraphics[width=\textwidth]{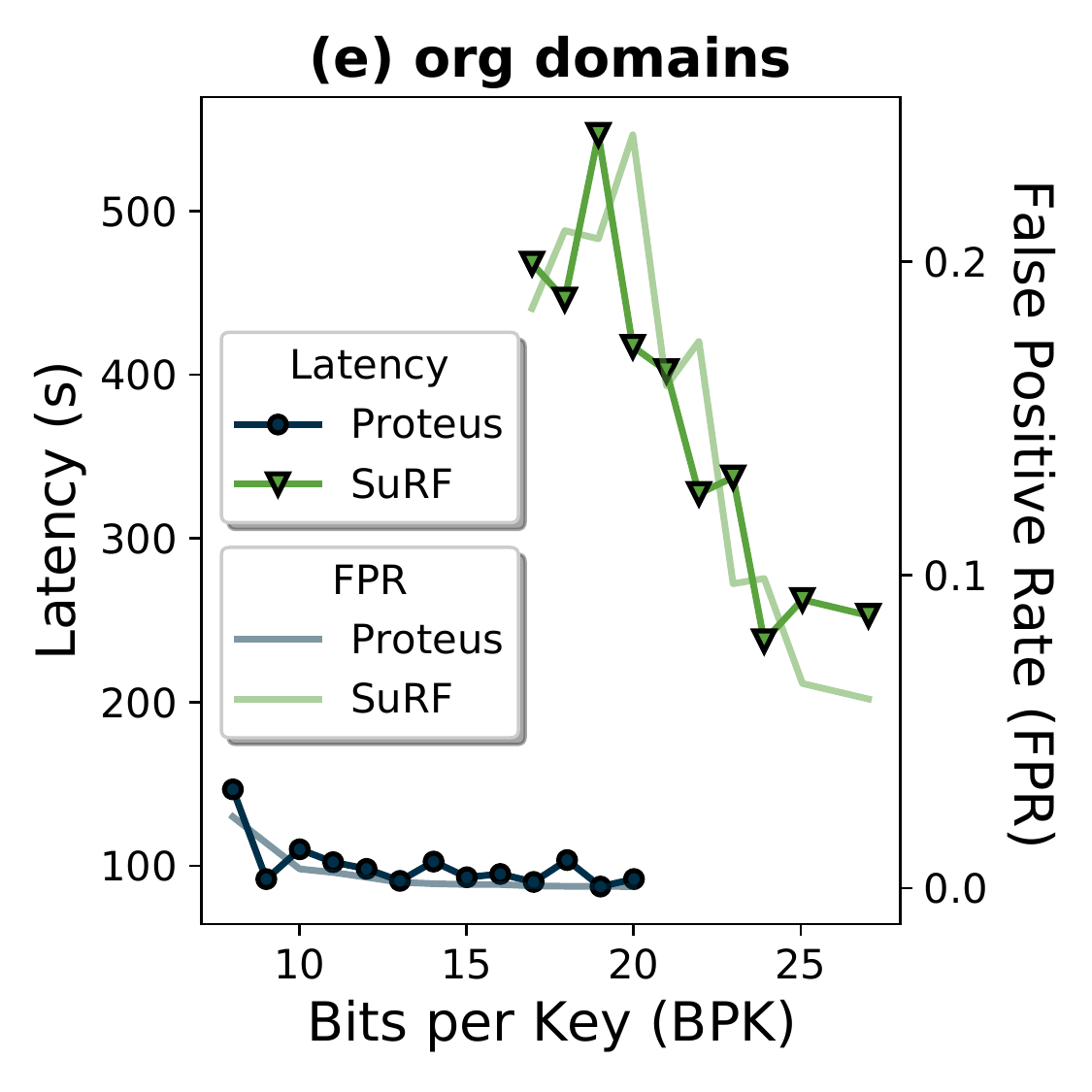}
     \end{subfigure}
    \vspace{-1em}
    \caption{Proteus achieves lower FPR than SuRF on synthetic strings (a-d) and lower RocksDB latency on real-world strings (e).}
    \label{fig:strings}
    \vspace{3pt}
\end{figure*}

\noindent \textbf{FPR is Unaffected by Key Length:} In Figure \ref{fig:strings}a-d, we present in-memory results for 1440-bit keys.
As with integers, Proteus outperforms SuRF across distributions and filter sizes. 
Experiments with other key lengths show the same performance patterns as only the range size and key-query proximity matter.

\noindent \textbf{Modeling Time Increases with Key Length:} While the modeling accuracy is preserved across key length, the number of possible designs is $\mathcal{O}(k^2)$ with respect to the maximum key length.
Longer keys can therefore result in a significant increase in the time required to model the designs.
In our in-memory benchmarks, the time to model all possible designs with 1440-bit keys ranged from 2.82 to 149 seconds.
The worst case is too expensive for most real world applications; however, we can achieve an order of magnitude speedup by using a coarser search as the difference in performance for similar designs is often quite small.
The results for Proteus shown in Figure \ref{fig:strings} were obtained by only modeling 128 uniformly spaced Bloom filter prefix lengths for all feasible trie depths. 
The modeling times with this optimization ranged from 0.86 to 14.3 seconds while still achieving similar performance.  

The structure of the modeling also lends itself very well to parallel computation. 
As shown in Section \ref{sec:cpfprmodel}, the dominant cost of the sampling is extracting the information needed to model each sample query relative to each design.  
This can be done independently on both a per-query and per-design basis. 
Since modern database services are hosted on elastic cloud architectures, occasional increases in CPU usage can easily be amortized at a low cost compared to the benefits of a more performant filter.

\noindent \textbf{Proteus Maintains Strong Performance on Strings:}
Figure \ref{fig:strings} shows the results of our real-world string benchmark in RocksDB.  
Proteus outperforms SuRF in both end-to-end latency and FPR by an even larger margin using the aforementioned coarse-grained modeling.
We see the impact of design tradeoffs amplified for longer keys \textemdash SuRF's rigid design requires a large minimum memory budget and limits the effectiveness of additional filter memory.
Conversely, Proteus's flexible design allows it to distribute memory between its design elements for more efficient memory use.
As shown in Figure \ref{fig:strings}, SuRF requires at least 16 BPK while Proteus can achieve significant performance gains with as little as 8 BPK.

\vspace{-0.6em}
\section{Related Work}
Our investigation is part of a broader initiative in the systems community to design contextually-customized data structures. 

\noindent \textbf{Adaptive Range Filter:} The Adaptive Range Filter (ARF) \cite{siberia} adapts its binary trie structure in response to queries, extending branches to compensate for false positives and retracting them to maintain its memory footprint.
However, ARF's encoding strategy limits its memory efficiency and requires significant time and memory to pre-train \cite{surf}. 
Similar adaptive techniques have been applied to AMQs to deal with adversarial workloads \cite{adaptive_cuckoo, broom_filter}.

\noindent \textbf{Stacked Filters:} Stacked Filters \cite{stacked_filters} also use modeling to incorporate workload specific information in their design. 
However, their model is designed for point rather than range queries. 

\noindent \textbf{Data Calculator:} Similar to PRFs, the Data Calculator \cite{data_calculator} breaks down a complex design space into its fundamental design primitives and models the behavior of designs to determine the optimal design for a given workload. However, the Data Calculator focuses on the design space of key-value structures rather than range filters and merely synthesizes the optimal design rather than instantiating it.

\noindent \textbf{Learned Structures:} Several iterations of learned Bloom filters use learned models to leverage patterns in the data for better performance \cite{lbf_model, plbf, albf}. 
These filters are designed for single item queries and perform poorly on range queries for the same reasons as other AMQs.  
This method of fitting to the use case also requires the presence of patterns in the data that are amenable to the model being used. 
Similar techniques have been applied to indexing structures in databases to speed up searches \cite{learned_index_structures}. 
Indexes can also be used to answer range queries, but they are larger, general purpose structures which typically require more I/O to answer a query.  
\vspace{-0.5em}
\section{Summary and Opportunities}
This paper introduces Proteus, a self-designing range filter that achieves robust performance across a large variety of workloads.
The core idea is that (1) Proteus unifies the design spaces of state-of-the-art range filters to cover a wider range of workloads and (2) is able to instantiate workload-optimal designs.
Analysing cutting-edge range filtering techniques through the lens of the CPFPR model reveals adversarial workloads which no current design can handle practically, suggesting the need for further expansion of the range filter design space. 
Other promising directions include extending the CPFPR model to support higher order optimization strategies by incorporating metrics such as query latency as well as exploring non-uniform memory allocation strategies in RocksDB.

\begin{acks}
This work is supported in part by NSF grants CCF-2101140, CNS-2107078, CCF-1563710, and DMS-2023528, and by a gift to the Center for Research on Computation and Society at Harvard University.
Additional funding was provided by USA Department of Energy project DE-SC0020200.
We would also like to thank our reviewers for their helpful and constructive feedback. 
\end{acks}

\bibliographystyle{ACM-Reference-Format}
\bibliography{works_cited}


\begin{thebibliography}{48}


\ifx \showCODEN    \undefined \def \showCODEN     #1{\unskip}     \fi
\ifx \showDOI      \undefined \def \showDOI       #1{#1}\fi
\ifx \showISBNx    \undefined \def \showISBNx     #1{\unskip}     \fi
\ifx \showISBNxiii \undefined \def \showISBNxiii  #1{\unskip}     \fi
\ifx \showISSN     \undefined \def \showISSN      #1{\unskip}     \fi
\ifx \showLCCN     \undefined \def \showLCCN      #1{\unskip}     \fi
\ifx \shownote     \undefined \def \shownote      #1{#1}          \fi
\ifx \showarticletitle \undefined \def \showarticletitle #1{#1}   \fi
\ifx \showURL      \undefined \def \showURL       {\relax}        \fi
\providecommand\bibfield[2]{#2}
\providecommand\bibinfo[2]{#2}
\providecommand\natexlab[1]{#1}
\providecommand\showeprint[2][]{arXiv:#2}

\bibitem[\protect\citeauthoryear{Alexiou, Kossmann, and Larson}{Alexiou
  et~al\mbox{.}}{2013}]%
        {siberia}
\bibfield{author}{\bibinfo{person}{Karolina Alexiou}, \bibinfo{person}{Donald
  Kossmann}, {and} \bibinfo{person}{Paul Larson}.}
  \bibinfo{year}{2013}\natexlab{}.
\newblock \showarticletitle{Adaptive Range Filters for Cold Data: Avoiding
  Trips to Siberia}. In \bibinfo{booktitle}{\emph{Proceedings of the VLDB
  Endowment, Vol. 6, No. 14}}.
\newblock
\urldef\tempurl%
\url{https://www.microsoft.com/en-us/research/publication/adaptive-range-filters-for-cold-data-avoiding-trips-to-siberia/}
\showURL{%
\tempurl}


\bibitem[\protect\citeauthoryear{Alsubaiee, Altowim, Altwaijry, Behm, Borkar,
  Bu, Carey, Cetindil, Cheelangi, Faraaz, et~al\mbox{.}}{Alsubaiee
  et~al\mbox{.}}{2014}]%
        {alsubaiee2014asterixdb}
\bibfield{author}{\bibinfo{person}{Sattam Alsubaiee}, \bibinfo{person}{Yasser
  Altowim}, \bibinfo{person}{Hotham Altwaijry}, \bibinfo{person}{Alexander
  Behm}, \bibinfo{person}{Vinayak Borkar}, \bibinfo{person}{Yingyi Bu},
  \bibinfo{person}{Michael Carey}, \bibinfo{person}{Inci Cetindil},
  \bibinfo{person}{Madhusudan Cheelangi}, \bibinfo{person}{Khurram Faraaz},
  {et~al\mbox{.}}} \bibinfo{year}{2014}\natexlab{}.
\newblock \showarticletitle{AsterixDB: A scalable, open source BDMS}.
\newblock \bibinfo{journal}{\emph{arXiv preprint arXiv:1407.0454}}
  (\bibinfo{year}{2014}).
\newblock


\bibitem[\protect\citeauthoryear{Alsubaiee, Carey, and Li}{Alsubaiee
  et~al\mbox{.}}{2015}]%
        {gisdatabase}
\bibfield{author}{\bibinfo{person}{Sattam Alsubaiee},
  \bibinfo{person}{Michael~J. Carey}, {and} \bibinfo{person}{Chen Li}.}
  \bibinfo{year}{2015}\natexlab{}.
\newblock \showarticletitle{LSM-Based Storage and Indexing: An Old Idea with
  Timely Benefits}. In \bibinfo{booktitle}{\emph{Second International ACM
  Workshop on Managing and Mining Enriched Geo-Spatial Data}} (Melbourne, VIC,
  Australia) \emph{(\bibinfo{series}{GeoRich'15})}.
  \bibinfo{publisher}{Association for Computing Machinery},
  \bibinfo{address}{New York, NY, USA}, \bibinfo{pages}{1–6}.
\newblock
\showISBNx{9781450336680}
\urldef\tempurl%
\url{https://doi.org/10.1145/2786006.2786007}
\showDOI{\tempurl}


\bibitem[\protect\citeauthoryear{Amrouche, Basara, Calafiura, Emeliyanov,
  Estrade, Farrell, Germain, Gligorov, Golling, Gorbunov, Gray, Guyon,
  Hushchyn, Innocente, Kiehn, Kunze, Moyse, Rousseau, Salzburger, Ustyuzhanin,
  and Vlimant}{Amrouche et~al\mbox{.}}{2021}]%
        {amrouche2021tracking}
\bibfield{author}{\bibinfo{person}{Sabrina Amrouche}, \bibinfo{person}{Laurent
  Basara}, \bibinfo{person}{Paolo Calafiura}, \bibinfo{person}{Dmitry
  Emeliyanov}, \bibinfo{person}{Victor Estrade}, \bibinfo{person}{Steven
  Farrell}, \bibinfo{person}{Cécile Germain}, \bibinfo{person}{Vladimir~Vava
  Gligorov}, \bibinfo{person}{Tobias Golling}, \bibinfo{person}{Sergey
  Gorbunov}, \bibinfo{person}{Heather Gray}, \bibinfo{person}{Isabelle Guyon},
  \bibinfo{person}{Mikhail Hushchyn}, \bibinfo{person}{Vincenzo Innocente},
  \bibinfo{person}{Moritz Kiehn}, \bibinfo{person}{Marcel Kunze},
  \bibinfo{person}{Edward Moyse}, \bibinfo{person}{David Rousseau},
  \bibinfo{person}{Andreas Salzburger}, \bibinfo{person}{Andrey Ustyuzhanin},
  {and} \bibinfo{person}{Jean-Roch Vlimant}.} \bibinfo{year}{2021}\natexlab{}.
\newblock \bibinfo{title}{The Tracking Machine Learning challenge : Throughput
  phase}.
\newblock
\newblock
\showeprint[arxiv]{2105.01160}~[cs.LG]


\bibitem[\protect\citeauthoryear{Appleby}{Appleby}{2008}]%
        {murmurhash}
\bibfield{author}{\bibinfo{person}{Austin Appleby}.}
  \bibinfo{year}{2008}\natexlab{}.
\newblock \bibinfo{booktitle}{}.
\newblock
\urldef\tempurl%
\url{https://sites.google.com/site/murmurhash/}
\showURL{%
\tempurl}


\bibitem[\protect\citeauthoryear{Arroyuelo, C\'{a}novas, Navarro, and
  Sadakane}{Arroyuelo et~al\mbox{.}}{2010}]%
        {succinct_trees}
\bibfield{author}{\bibinfo{person}{Diego Arroyuelo}, \bibinfo{person}{Rodrigo
  C\'{a}novas}, \bibinfo{person}{Gonzalo Navarro}, {and}
  \bibinfo{person}{Kunihiko Sadakane}.} \bibinfo{year}{2010}\natexlab{}.
\newblock \showarticletitle{Succinct Trees in Practice}. In
  \bibinfo{booktitle}{\emph{Proceedings of the Meeting on Algorithm Engineering
  \& Expermiments}} (Austin, Texas) \emph{(\bibinfo{series}{ALENEX '10})}.
  \bibinfo{publisher}{Society for Industrial and Applied Mathematics},
  \bibinfo{address}{USA}, \bibinfo{pages}{84–97}.
\newblock


\bibitem[\protect\citeauthoryear{Bender, Farach-Colton, Goswami, Johnson,
  McCauley, and Singh}{Bender et~al\mbox{.}}{2018}]%
        {broom_filter}
\bibfield{author}{\bibinfo{person}{Michael~A. Bender}, \bibinfo{person}{Martin
  Farach-Colton}, \bibinfo{person}{Mayank Goswami}, \bibinfo{person}{Rob
  Johnson}, \bibinfo{person}{Samuel McCauley}, {and} \bibinfo{person}{Shikha
  Singh}.} \bibinfo{year}{2018}\natexlab{}.
\newblock \showarticletitle{Bloom Filters, Adaptivity, and the Dictionary
  Problem}. In \bibinfo{booktitle}{\emph{2018 IEEE 59th Annual Symposium on
  Foundations of Computer Science (FOCS)}}. \bibinfo{pages}{182--193}.
\newblock
\urldef\tempurl%
\url{https://doi.org/10.1109/FOCS.2018.00026}
\showDOI{\tempurl}


\bibitem[\protect\citeauthoryear{Bender, Farach-Colton, Johnson, Kraner,
  Kuszmaul, Medjedovic, Montes, Shetty, Spillane, and Zadok}{Bender
  et~al\mbox{.}}{2012}]%
        {quotient_filter}
\bibfield{author}{\bibinfo{person}{Michael~A. Bender}, \bibinfo{person}{Martin
  Farach-Colton}, \bibinfo{person}{Rob Johnson}, \bibinfo{person}{Russell
  Kraner}, \bibinfo{person}{Bradley~C. Kuszmaul}, \bibinfo{person}{Dzejla
  Medjedovic}, \bibinfo{person}{Pablo Montes}, \bibinfo{person}{Pradeep
  Shetty}, \bibinfo{person}{Richard~P. Spillane}, {and} \bibinfo{person}{Erez
  Zadok}.} \bibinfo{year}{2012}\natexlab{}.
\newblock \showarticletitle{Don't Thrash: How to Cache Your Hash on Flash}.
\newblock \bibinfo{journal}{\emph{Proc. VLDB Endow.}} \bibinfo{volume}{5},
  \bibinfo{number}{11} (\bibinfo{date}{July} \bibinfo{year}{2012}),
  \bibinfo{pages}{1627–1637}.
\newblock
\showISSN{2150-8097}
\urldef\tempurl%
\url{https://doi.org/10.14778/2350229.2350275}
\showDOI{\tempurl}


\bibitem[\protect\citeauthoryear{Benoit, Demaine, Munro, Raman, Raman, and
  Rao}{Benoit et~al\mbox{.}}{2005}]%
        {trees_higher_degree}
\bibfield{author}{\bibinfo{person}{David Benoit}, \bibinfo{person}{Erik~D.
  Demaine}, \bibinfo{person}{J.~Ian Munro}, \bibinfo{person}{Rajeev Raman},
  \bibinfo{person}{Venkatesh Raman}, {and} \bibinfo{person}{S.~Srinivasa Rao}.}
  \bibinfo{year}{2005}\natexlab{}.
\newblock \showarticletitle{Representing Trees of Higher Degree}.
\newblock \bibinfo{journal}{\emph{Algorithmica}} \bibinfo{volume}{43},
  \bibinfo{number}{4} (\bibinfo{date}{Dec.} \bibinfo{year}{2005}),
  \bibinfo{pages}{275–292}.
\newblock
\showISSN{0178-4617}


\bibitem[\protect\citeauthoryear{Bloom}{Bloom}{1970}]%
        {bloom_filter}
\bibfield{author}{\bibinfo{person}{Burton~H. Bloom}.}
  \bibinfo{year}{1970}\natexlab{}.
\newblock \showarticletitle{Space/Time Trade-Offs in Hash Coding with Allowable
  Errors}.
\newblock \bibinfo{journal}{\emph{Commun. ACM}} \bibinfo{volume}{13},
  \bibinfo{number}{7} (\bibinfo{date}{July} \bibinfo{year}{1970}),
  \bibinfo{pages}{422–426}.
\newblock
\showISSN{0001-0782}
\urldef\tempurl%
\url{https://doi.org/10.1145/362686.362692}
\showDOI{\tempurl}


\bibitem[\protect\citeauthoryear{Bonomi, Mitzenmacher, Panigrahy, Singh, and
  Varghese}{Bonomi et~al\mbox{.}}{2006}]%
        {better_counting_bf}
\bibfield{author}{\bibinfo{person}{Flavio Bonomi}, \bibinfo{person}{Michael
  Mitzenmacher}, \bibinfo{person}{Rina Panigrahy}, \bibinfo{person}{Sushil
  Singh}, {and} \bibinfo{person}{George Varghese}.}
  \bibinfo{year}{2006}\natexlab{}.
\newblock \showarticletitle{An Improved Construction for Counting Bloom
  Filters} \emph{(\bibinfo{series}{ESA'06})}.
  \bibinfo{publisher}{Springer-Verlag}, \bibinfo{address}{Berlin, Heidelberg},
  \bibinfo{pages}{684–695}.
\newblock
\showISBNx{3540388753}
\urldef\tempurl%
\url{https://doi.org/10.1007/11841036_61}
\showDOI{\tempurl}


\bibitem[\protect\citeauthoryear{Cao, Dong, Vemuri, and Du}{Cao
  et~al\mbox{.}}{2020}]%
        {facebook_workloads}
\bibfield{author}{\bibinfo{person}{Zhichao Cao}, \bibinfo{person}{Siying Dong},
  \bibinfo{person}{Sagar Vemuri}, {and} \bibinfo{person}{David~H.C. Du}.}
  \bibinfo{year}{2020}\natexlab{}.
\newblock \showarticletitle{Characterizing, Modeling, and Benchmarking RocksDB
  Key-Value Workloads at Facebook}. In \bibinfo{booktitle}{\emph{18th {USENIX}
  Conference on File and Storage Technologies ({FAST} 20)}}.
  \bibinfo{publisher}{{USENIX} Association}, \bibinfo{address}{Santa Clara,
  CA}, \bibinfo{pages}{209--223}.
\newblock
\showISBNx{978-1-939133-12-0}
\urldef\tempurl%
\url{https://www.usenix.org/conference/fast20/presentation/cao-zhichao}
\showURL{%
\tempurl}


\bibitem[\protect\citeauthoryear{Chang, Dean, Ghemawat, Hsieh, Wallach,
  Burrows, Chandra, Fikes, and Gruber}{Chang et~al\mbox{.}}{2008}]%
        {chang2008bigtable}
\bibfield{author}{\bibinfo{person}{Fay Chang}, \bibinfo{person}{Jeffrey Dean},
  \bibinfo{person}{Sanjay Ghemawat}, \bibinfo{person}{Wilson~C Hsieh},
  \bibinfo{person}{Deborah~A Wallach}, \bibinfo{person}{Mike Burrows},
  \bibinfo{person}{Tushar Chandra}, \bibinfo{person}{Andrew Fikes}, {and}
  \bibinfo{person}{Robert~E Gruber}.} \bibinfo{year}{2008}\natexlab{}.
\newblock \showarticletitle{Bigtable: A distributed storage system for
  structured data}.
\newblock \bibinfo{journal}{\emph{ACM Transactions on Computer Systems (TOCS)}}
  \bibinfo{volume}{26}, \bibinfo{number}{2} (\bibinfo{year}{2008}),
  \bibinfo{pages}{1--26}.
\newblock


\bibitem[\protect\citeauthoryear{Dai and Shrivastava}{Dai and
  Shrivastava}{2019}]%
        {albf}
\bibfield{author}{\bibinfo{person}{Zhenwei Dai} {and}
  \bibinfo{person}{Anshumali Shrivastava}.} \bibinfo{year}{2019}\natexlab{}.
\newblock \showarticletitle{Adaptive Learned Bloom Filter (Ada-BF): Efficient
  Utilization of the Classifier}.
\newblock \bibinfo{journal}{\emph{CoRR}}  \bibinfo{volume}{abs/1910.09131}
  (\bibinfo{year}{2019}).
\newblock
\showeprint[arXiv]{1910.09131}
\urldef\tempurl%
\url{http://arxiv.org/abs/1910.09131}
\showURL{%
\tempurl}


\bibitem[\protect\citeauthoryear{Deeds, Hentschel, and Idreos}{Deeds
  et~al\mbox{.}}{2021}]%
        {stacked_filters}
\bibfield{author}{\bibinfo{person}{Kyle Deeds}, \bibinfo{person}{Brian
  Hentschel}, {and} \bibinfo{person}{Stratos Idreos}.}
  \bibinfo{year}{2021}\natexlab{}.
\newblock \showarticletitle{Stacked Filters: Learning to Filter by Structure}.
\newblock \bibinfo{journal}{\emph{Proceedings of the VLDB Endowment}}
  \bibinfo{volume}{14}, \bibinfo{number}{4} (\bibinfo{year}{2021}),
  \bibinfo{pages}{600 -- 612}.
\newblock


\bibitem[\protect\citeauthoryear{Dgraph.}{Dgraph.}{2017}]%
        {dgraph}
\bibfield{author}{\bibinfo{person}{Dgraph.}} \bibinfo{year}{2017}\natexlab{}.
\newblock \bibinfo{booktitle}{\emph{Fast Key-value DB in Go.}}
\newblock
\urldef\tempurl%
\url{https://github.com/dgraph-io/badger}
\showURL{%
\tempurl}


\bibitem[\protect\citeauthoryear{Dharmapurikar, Krishnamurthy, and
  Taylor}{Dharmapurikar et~al\mbox{.}}{2006}]%
        {prefix_matching}
\bibfield{author}{\bibinfo{person}{S. Dharmapurikar}, \bibinfo{person}{P.
  Krishnamurthy}, {and} \bibinfo{person}{D.E. Taylor}.}
  \bibinfo{year}{2006}\natexlab{}.
\newblock \showarticletitle{Longest prefix matching using bloom filters}.
\newblock \bibinfo{journal}{\emph{IEEE/ACM Transactions on Networking}}
  \bibinfo{volume}{14}, \bibinfo{number}{2} (\bibinfo{year}{2006}),
  \bibinfo{pages}{397--409}.
\newblock
\urldef\tempurl%
\url{https://doi.org/10.1109/TNET.2006.872576}
\showDOI{\tempurl}


\bibitem[\protect\citeauthoryear{Dillinger and Walzer}{Dillinger and
  Walzer}{2021}]%
        {ribbon_filter}
\bibfield{author}{\bibinfo{person}{Peter~C. Dillinger} {and}
  \bibinfo{person}{Stefan Walzer}.} \bibinfo{year}{2021}\natexlab{}.
\newblock \showarticletitle{Ribbon filter: practically smaller than Bloom and
  Xor}.
\newblock \bibinfo{journal}{\emph{CoRR}}  \bibinfo{volume}{abs/2103.02515}
  (\bibinfo{year}{2021}).
\newblock
\showeprint[arxiv]{2103.02515}
\urldef\tempurl%
\url{https://arxiv.org/abs/2103.02515}
\showURL{%
\tempurl}


\bibitem[\protect\citeauthoryear{Dong, Kryczka, Jin, and Stumm}{Dong
  et~al\mbox{.}}{2021}]%
        {dong2021evolution}
\bibfield{author}{\bibinfo{person}{Siying Dong}, \bibinfo{person}{Andrew
  Kryczka}, \bibinfo{person}{Yanqin Jin}, {and} \bibinfo{person}{Michael
  Stumm}.} \bibinfo{year}{2021}\natexlab{}.
\newblock \showarticletitle{Evolution of Development Priorities in Key-value
  Stores Serving Large-scale Applications: The $\{$RocksDB$\}$ Experience}. In
  \bibinfo{booktitle}{\emph{19th USENIX Conference on File and Storage
  Technologies (FAST 21)}}. \bibinfo{pages}{33--49}.
\newblock


\bibitem[\protect\citeauthoryear{Fan, Andersen, Kaminsky, and Mitzenmacher}{Fan
  et~al\mbox{.}}{2014}]%
        {cuckoo_filter}
\bibfield{author}{\bibinfo{person}{Bin Fan}, \bibinfo{person}{Dave~G.
  Andersen}, \bibinfo{person}{Michael Kaminsky}, {and}
  \bibinfo{person}{Michael~D. Mitzenmacher}.} \bibinfo{year}{2014}\natexlab{}.
\newblock \showarticletitle{Cuckoo Filter: Practically Better Than Bloom}. In
  \bibinfo{booktitle}{\emph{Proceedings of the 10th ACM International on
  Conference on Emerging Networking Experiments and Technologies}} (Sydney,
  Australia) \emph{(\bibinfo{series}{CoNEXT '14})}.
  \bibinfo{publisher}{Association for Computing Machinery},
  \bibinfo{address}{New York, NY, USA}, \bibinfo{pages}{75–88}.
\newblock
\showISBNx{9781450332798}
\urldef\tempurl%
\url{https://doi.org/10.1145/2674005.2674994}
\showDOI{\tempurl}


\bibitem[\protect\citeauthoryear{Fredkin}{Fredkin}{1960}]%
        {trie_memory}
\bibfield{author}{\bibinfo{person}{Edward Fredkin}.}
  \bibinfo{year}{1960}\natexlab{}.
\newblock \showarticletitle{Trie Memory}.
\newblock \bibinfo{journal}{\emph{Commun. ACM}} \bibinfo{volume}{3},
  \bibinfo{number}{9} (\bibinfo{date}{Sept.} \bibinfo{year}{1960}),
  \bibinfo{pages}{490–499}.
\newblock
\showISSN{0001-0782}
\urldef\tempurl%
\url{https://doi.org/10.1145/367390.367400}
\showDOI{\tempurl}


\bibitem[\protect\citeauthoryear{Ge, Li, Yuan, and Huang}{Ge
  et~al\mbox{.}}{2019}]%
        {skewed_range_query}
\bibfield{author}{\bibinfo{person}{Wei Ge}, \bibinfo{person}{Xianxian Li},
  \bibinfo{person}{Chunfeng Yuan}, {and} \bibinfo{person}{Yihua Huang}.}
  \bibinfo{year}{2019}\natexlab{}.
\newblock \showarticletitle{Correlation-Aware Partitioning for Skewed Range
  Query Optimization}.
\newblock \bibinfo{journal}{\emph{World Wide Web}} \bibinfo{volume}{22},
  \bibinfo{number}{1} (\bibinfo{date}{Jan.} \bibinfo{year}{2019}),
  \bibinfo{pages}{125–151}.
\newblock
\showISSN{1386-145X}
\urldef\tempurl%
\url{https://doi.org/10.1007/s11280-018-0547-4}
\showDOI{\tempurl}


\bibitem[\protect\citeauthoryear{Goswami, Grønlund, Larsen, and Pagh}{Goswami
  et~al\mbox{.}}{[n.d.]}]%
        {optimal_space}
\bibfield{author}{\bibinfo{person}{Mayank Goswami}, \bibinfo{person}{Allan
  Grønlund}, \bibinfo{person}{Kasper~Green Larsen}, {and}
  \bibinfo{person}{Rasmus Pagh}.} \bibinfo{year}{[n.d.]}\natexlab{}.
\newblock \bibinfo{booktitle}{\emph{Approximate Range Emptiness in Constant
  Time and Optimal Space}}.
\newblock \bibinfo{pages}{769--775}.
\newblock
\urldef\tempurl%
\url{https://doi.org/10.1137/1.9781611973730.52}
\showDOI{\tempurl}
\showeprint{https://epubs.siam.org/doi/pdf/10.1137/1.9781611973730.52}


\bibitem[\protect\citeauthoryear{Graf and Lemire}{Graf and Lemire}{2019}]%
        {xor_filter}
\bibfield{author}{\bibinfo{person}{Thomas~Mueller Graf} {and}
  \bibinfo{person}{Daniel Lemire}.} \bibinfo{year}{2019}\natexlab{}.
\newblock \showarticletitle{Xor Filters: Faster and Smaller Than Bloom and
  Cuckoo Filters}.
\newblock \bibinfo{journal}{\emph{CoRR}}  \bibinfo{volume}{abs/1912.08258}
  (\bibinfo{year}{2019}).
\newblock
\showeprint[arxiv]{1912.08258}
\urldef\tempurl%
\url{http://arxiv.org/abs/1912.08258}
\showURL{%
\tempurl}


\bibitem[\protect\citeauthoryear{Idreos and Callaghan}{Idreos and
  Callaghan}{2020}]%
        {idreos2020key}
\bibfield{author}{\bibinfo{person}{Stratos Idreos} {and} \bibinfo{person}{Mark
  Callaghan}.} \bibinfo{year}{2020}\natexlab{}.
\newblock \showarticletitle{Key-value storage engines}. In
  \bibinfo{booktitle}{\emph{Proceedings of the 2020 ACM SIGMOD International
  Conference on Management of Data}}. \bibinfo{pages}{2667--2672}.
\newblock


\bibitem[\protect\citeauthoryear{Idreos, Zoumpatianos, Hentschel, Kester, and
  Guo}{Idreos et~al\mbox{.}}{2018}]%
        {data_calculator}
\bibfield{author}{\bibinfo{person}{Stratos Idreos}, \bibinfo{person}{Kostas
  Zoumpatianos}, \bibinfo{person}{Brian Hentschel}, \bibinfo{person}{Michael~S.
  Kester}, {and} \bibinfo{person}{Demi Guo}.} \bibinfo{year}{2018}\natexlab{}.
\newblock \showarticletitle{The Data Calculator: Data Structure Design and Cost
  Synthesis from First Principles and Learned Cost Models}. In
  \bibinfo{booktitle}{\emph{Proceedings of the 2018 International Conference on
  Management of Data}} (Houston, TX, USA) \emph{(\bibinfo{series}{SIGMOD
  '18})}. \bibinfo{publisher}{Association for Computing Machinery},
  \bibinfo{address}{New York, NY, USA}, \bibinfo{pages}{535–550}.
\newblock
\showISBNx{9781450347037}
\urldef\tempurl%
\url{https://doi.org/10.1145/3183713.3199671}
\showDOI{\tempurl}


\bibitem[\protect\citeauthoryear{Inc.}{Inc.}{2012}]%
        {rocksdb}
\bibfield{author}{\bibinfo{person}{Facebook Inc.}}
  \bibinfo{year}{2012}\natexlab{}.
\newblock \bibinfo{booktitle}{\emph{RocksDB}}.
\newblock
\urldef\tempurl%
\url{https://github.com/facebook/rocksdb}
\showURL{%
\tempurl}


\bibitem[\protect\citeauthoryear{Inc.}{Inc.}{2020}]%
        {rocksdb_compression}
\bibfield{author}{\bibinfo{person}{Facebook Inc.}}
  \bibinfo{year}{2020}\natexlab{}.
\newblock \bibinfo{booktitle}{\emph{Compression}}.
\newblock
\urldef\tempurl%
\url{https://github.com/facebook/rocksdb/wiki/Compression}
\showURL{%
\tempurl}


\bibitem[\protect\citeauthoryear{Jacobson}{Jacobson}{1989}]%
        {efficient_trees}
\bibfield{author}{\bibinfo{person}{G. Jacobson}.}
  \bibinfo{year}{1989}\natexlab{}.
\newblock \showarticletitle{Space-efficient static trees and graphs}. In
  \bibinfo{booktitle}{\emph{30th Annual Symposium on Foundations of Computer
  Science}}. \bibinfo{pages}{549--554}.
\newblock
\urldef\tempurl%
\url{https://doi.org/10.1109/SFCS.1989.63533}
\showDOI{\tempurl}


\bibitem[\protect\citeauthoryear{Kipf, Marcus, van Renen, Stoian, Kemper,
  Kraska, and Neumann}{Kipf et~al\mbox{.}}{2019}]%
        {sosd-neurips}
\bibfield{author}{\bibinfo{person}{Andreas Kipf}, \bibinfo{person}{Ryan
  Marcus}, \bibinfo{person}{Alexander van Renen}, \bibinfo{person}{Mihail
  Stoian}, \bibinfo{person}{Alfons Kemper}, \bibinfo{person}{Tim Kraska}, {and}
  \bibinfo{person}{Thomas Neumann}.} \bibinfo{year}{2019}\natexlab{}.
\newblock \showarticletitle{SOSD: A Benchmark for Learned Indexes}.
\newblock \bibinfo{journal}{\emph{NeurIPS Workshop on Machine Learning for
  Systems}} (\bibinfo{year}{2019}).
\newblock


\bibitem[\protect\citeauthoryear{Kondylakis, Dayan, Zoumpatianos, and
  Palpanas}{Kondylakis et~al\mbox{.}}{2020}]%
        {kondylakis2020coconut}
\bibfield{author}{\bibinfo{person}{Haridimos Kondylakis}, \bibinfo{person}{Niv
  Dayan}, \bibinfo{person}{Kostas Zoumpatianos}, {and} \bibinfo{person}{Themis
  Palpanas}.} \bibinfo{year}{2020}\natexlab{}.
\newblock \bibinfo{title}{Coconut Palm: Static and Streaming Data Series
  Exploration Now in your Palm}.
\newblock
\newblock
\showeprint[arxiv]{2006.13079}~[cs.DB]


\bibitem[\protect\citeauthoryear{Kraska, Beutel, Chi, Dean, and
  Polyzotis}{Kraska et~al\mbox{.}}{2018}]%
        {learned_index_structures}
\bibfield{author}{\bibinfo{person}{Tim Kraska}, \bibinfo{person}{Alex Beutel},
  \bibinfo{person}{Ed~H. Chi}, \bibinfo{person}{Jeffrey Dean}, {and}
  \bibinfo{person}{Neoklis Polyzotis}.} \bibinfo{year}{2018}\natexlab{}.
\newblock \showarticletitle{The Case for Learned Index Structures}. In
  \bibinfo{booktitle}{\emph{Proceedings of the 2018 International Conference on
  Management of Data}} (Houston, TX, USA) \emph{(\bibinfo{series}{SIGMOD
  '18})}. \bibinfo{publisher}{Association for Computing Machinery},
  \bibinfo{address}{New York, NY, USA}, \bibinfo{pages}{489–504}.
\newblock
\showISBNx{9781450347037}
\urldef\tempurl%
\url{https://doi.org/10.1145/3183713.3196909}
\showDOI{\tempurl}


\bibitem[\protect\citeauthoryear{Kyrola and Guestrin}{Kyrola and
  Guestrin}{2014}]%
        {kyrola2014graphchidb}
\bibfield{author}{\bibinfo{person}{Aapo Kyrola} {and} \bibinfo{person}{Carlos
  Guestrin}.} \bibinfo{year}{2014}\natexlab{}.
\newblock \bibinfo{title}{GraphChi-DB: Simple Design for a Scalable Graph
  Database System -- on Just a PC}.
\newblock
\newblock
\showeprint[arxiv]{1403.0701}~[cs.DB]


\bibitem[\protect\citeauthoryear{Lakshman and Malik}{Lakshman and
  Malik}{2010}]%
        {lakshman2010cassandra}
\bibfield{author}{\bibinfo{person}{Avinash Lakshman} {and}
  \bibinfo{person}{Prashant Malik}.} \bibinfo{year}{2010}\natexlab{}.
\newblock \showarticletitle{Cassandra: a decentralized structured storage
  system}.
\newblock \bibinfo{journal}{\emph{ACM SIGOPS Operating Systems Review}}
  \bibinfo{volume}{44}, \bibinfo{number}{2} (\bibinfo{year}{2010}),
  \bibinfo{pages}{35--40}.
\newblock


\bibitem[\protect\citeauthoryear{Lemire and Kaser}{Lemire and Kaser}{2016}]%
        {lemire2016faster}
\bibfield{author}{\bibinfo{person}{Daniel Lemire} {and} \bibinfo{person}{Owen
  Kaser}.} \bibinfo{year}{2016}\natexlab{}.
\newblock \showarticletitle{Faster 64-bit universal hashing using carry-less
  multiplications}.
\newblock \bibinfo{journal}{\emph{Journal of Cryptographic Engineering}}
  \bibinfo{volume}{6}, \bibinfo{number}{3} (\bibinfo{year}{2016}),
  \bibinfo{pages}{171--185}.
\newblock


\bibitem[\protect\citeauthoryear{Luo, Chatterjee, Ketsetsidis, Dayan, Qin, and
  Idreos}{Luo et~al\mbox{.}}{2020}]%
        {rosetta}
\bibfield{author}{\bibinfo{person}{Siqiang Luo}, \bibinfo{person}{Subarna
  Chatterjee}, \bibinfo{person}{Rafael Ketsetsidis}, \bibinfo{person}{Niv
  Dayan}, \bibinfo{person}{Wilson Qin}, {and} \bibinfo{person}{Stratos
  Idreos}.} \bibinfo{year}{2020}\natexlab{}.
\newblock \showarticletitle{Rosetta: A Robust Space-Time Optimized Range Filter
  for Key-Value Stores}. In \bibinfo{booktitle}{\emph{Proceedings of the 2020
  ACM SIGMOD International Conference on Management of Data}} (Portland, OR,
  USA) \emph{(\bibinfo{series}{SIGMOD '20})}. \bibinfo{publisher}{Association
  for Computing Machinery}, \bibinfo{address}{New York, NY, USA},
  \bibinfo{pages}{2071–2086}.
\newblock
\showISBNx{9781450367356}
\urldef\tempurl%
\url{https://doi.org/10.1145/3318464.3389731}
\showDOI{\tempurl}


\bibitem[\protect\citeauthoryear{Marcus, Kipf, van Renen, Stoian, Misra,
  Kemper, Neumann, and Kraska}{Marcus et~al\mbox{.}}{2020}]%
        {sosd-vldb}
\bibfield{author}{\bibinfo{person}{Ryan Marcus}, \bibinfo{person}{Andreas
  Kipf}, \bibinfo{person}{Alexander van Renen}, \bibinfo{person}{Mihail
  Stoian}, \bibinfo{person}{Sanchit Misra}, \bibinfo{person}{Alfons Kemper},
  \bibinfo{person}{Thomas Neumann}, {and} \bibinfo{person}{Tim Kraska}.}
  \bibinfo{year}{2020}\natexlab{}.
\newblock \showarticletitle{Benchmarking Learned Indexes}.
\newblock \bibinfo{journal}{\emph{Proc. {VLDB} Endow.}} \bibinfo{volume}{14},
  \bibinfo{number}{1} (\bibinfo{year}{2020}), \bibinfo{pages}{1--13}.
\newblock


\bibitem[\protect\citeauthoryear{Merriam-Webster}{Merriam-Webster}{[n.d.]}]%
        {mw:protean}
\bibfield{author}{\bibinfo{person}{Merriam-Webster}.}
  \bibinfo{year}{[n.d.]}\natexlab{}.
\newblock \showarticletitle{Protean}.
\newblock In \bibinfo{booktitle}{\emph{Merriam-Webster.com dictionary}}.
\newblock
\urldef\tempurl%
\url{https://www.merriam-webster.com/dictionary/Protean}
\showURL{%
\tempurl}


\bibitem[\protect\citeauthoryear{Mitzenmacher}{Mitzenmacher}{2018}]%
        {lbf_model}
\bibfield{author}{\bibinfo{person}{Michael Mitzenmacher}.}
  \bibinfo{year}{2018}\natexlab{}.
\newblock \showarticletitle{A Model for Learned Bloom Filters and Optimizing by
  Sandwiching}. In \bibinfo{booktitle}{\emph{Advances in Neural Information
  Processing Systems}}, \bibfield{editor}{\bibinfo{person}{S.~Bengio},
  \bibinfo{person}{H.~Wallach}, \bibinfo{person}{H.~Larochelle},
  \bibinfo{person}{K.~Grauman}, \bibinfo{person}{N.~Cesa-Bianchi}, {and}
  \bibinfo{person}{R.~Garnett}} (Eds.), Vol.~\bibinfo{volume}{31}.
  \bibinfo{publisher}{Curran Associates, Inc.}
\newblock
\urldef\tempurl%
\url{https://proceedings.neurips.cc/paper/2018/file/0f49c89d1e7298bb9930789c8ed59d48-Paper.pdf}
\showURL{%
\tempurl}


\bibitem[\protect\citeauthoryear{Mitzenmacher, Pontarelli, and
  Reviriego}{Mitzenmacher et~al\mbox{.}}{2017}]%
        {adaptive_cuckoo}
\bibfield{author}{\bibinfo{person}{Michael Mitzenmacher},
  \bibinfo{person}{Salvatore Pontarelli}, {and} \bibinfo{person}{Pedro
  Reviriego}.} \bibinfo{year}{2017}\natexlab{}.
\newblock \bibinfo{title}{Adaptive Cuckoo Filters}.
\newblock
\newblock
\showeprint[arxiv]{1704.06818}~[cs.DS]


\bibitem[\protect\citeauthoryear{Mitzenmacher and Upfal}{Mitzenmacher and
  Upfal}{2017}]%
        {mitzenmacher2017probability}
\bibfield{author}{\bibinfo{person}{Michael Mitzenmacher} {and}
  \bibinfo{person}{Eli Upfal}.} \bibinfo{year}{2017}\natexlab{}.
\newblock \bibinfo{booktitle}{\emph{Probability and computing: Randomization
  and probabilistic techniques in algorithms and data analysis}}.
\newblock \bibinfo{publisher}{Cambridge university press}.
\newblock


\bibitem[\protect\citeauthoryear{Pavlo, Curino, and Zdonik}{Pavlo
  et~al\mbox{.}}{2012}]%
        {skew_aware_db_partition}
\bibfield{author}{\bibinfo{person}{Andrew Pavlo}, \bibinfo{person}{Carlo
  Curino}, {and} \bibinfo{person}{Stanley Zdonik}.}
  \bibinfo{year}{2012}\natexlab{}.
\newblock \showarticletitle{Skew-Aware Automatic Database Partitioning in
  Shared-Nothing, Parallel OLTP Systems}. In
  \bibinfo{booktitle}{\emph{Proceedings of the 2012 ACM SIGMOD International
  Conference on Management of Data}} (Scottsdale, Arizona, USA)
  \emph{(\bibinfo{series}{SIGMOD '12})}. \bibinfo{publisher}{Association for
  Computing Machinery}, \bibinfo{address}{New York, NY, USA},
  \bibinfo{pages}{61–72}.
\newblock
\showISBNx{9781450312479}
\urldef\tempurl%
\url{https://doi.org/10.1145/2213836.2213844}
\showDOI{\tempurl}


\bibitem[\protect\citeauthoryear{Sivasubramanian}{Sivasubramanian}{2012}]%
        {sivasubramanian2012amazon}
\bibfield{author}{\bibinfo{person}{Swaminathan Sivasubramanian}.}
  \bibinfo{year}{2012}\natexlab{}.
\newblock \showarticletitle{Amazon dynamoDB: a seamlessly scalable
  non-relational database service}. In \bibinfo{booktitle}{\emph{Proceedings of
  the 2012 ACM SIGMOD International Conference on Management of Data}}.
  \bibinfo{pages}{729--730}.
\newblock


\bibitem[\protect\citeauthoryear{Turkynewych}{Turkynewych}{2022}]%
        {domains}
\bibfield{author}{\bibinfo{person}{Bohdan Turkynewych}.}
  \bibinfo{year}{2022}\natexlab{}.
\newblock \bibinfo{booktitle}{\emph{Domains Project}}.
\newblock
\urldef\tempurl%
\url{https://domainsproject.org/}
\showURL{%
Retrieved March 20, 2022 from \tempurl}


\bibitem[\protect\citeauthoryear{Vaidya, Knorr, Kraska, and
  Mitzenmacher}{Vaidya et~al\mbox{.}}{2020}]%
        {plbf}
\bibfield{author}{\bibinfo{person}{Kapil Vaidya}, \bibinfo{person}{Eric Knorr},
  \bibinfo{person}{Tim Kraska}, {and} \bibinfo{person}{Michael Mitzenmacher}.}
  \bibinfo{year}{2020}\natexlab{}.
\newblock \showarticletitle{Partitioned Learned Bloom Filter}.
\newblock \bibinfo{journal}{\emph{CoRR}}  \bibinfo{volume}{abs/2006.03176}
  (\bibinfo{year}{2020}).
\newblock
\showeprint[arXiv]{2006.03176}
\urldef\tempurl%
\url{https://arxiv.org/abs/2006.03176}
\showURL{%
\tempurl}


\bibitem[\protect\citeauthoryear{Xu, Zhang, and Xu}{Xu et~al\mbox{.}}{2019}]%
        {vchain}
\bibfield{author}{\bibinfo{person}{Cheng Xu}, \bibinfo{person}{Ce Zhang}, {and}
  \bibinfo{person}{Jianliang Xu}.} \bibinfo{year}{2019}\natexlab{}.
\newblock \showarticletitle{VChain: Enabling Verifiable Boolean Range Queries
  over Blockchain Databases}. In \bibinfo{booktitle}{\emph{Proceedings of the
  2019 International Conference on Management of Data}} (Amsterdam,
  Netherlands) \emph{(\bibinfo{series}{SIGMOD '19})}.
  \bibinfo{publisher}{Association for Computing Machinery},
  \bibinfo{address}{New York, NY, USA}, \bibinfo{pages}{141–158}.
\newblock
\showISBNx{9781450356435}
\urldef\tempurl%
\url{https://doi.org/10.1145/3299869.3300083}
\showDOI{\tempurl}


\bibitem[\protect\citeauthoryear{Zacharatou, \v{S}idlauskas, Tauheed, Heinis,
  and Ailamaki}{Zacharatou et~al\mbox{.}}{2019}]%
        {spatialrangequery}
\bibfield{author}{\bibinfo{person}{Eleni~Tzirita Zacharatou},
  \bibinfo{person}{Darius \v{S}idlauskas}, \bibinfo{person}{Farhan Tauheed},
  \bibinfo{person}{Thomas Heinis}, {and} \bibinfo{person}{Anastasia Ailamaki}.}
  \bibinfo{year}{2019}\natexlab{}.
\newblock \showarticletitle{Efficient Bundled Spatial Range Queries}. In
  \bibinfo{booktitle}{\emph{Proceedings of the 27th ACM SIGSPATIAL
  International Conference on Advances in Geographic Information Systems}}
  (Chicago, IL, USA) \emph{(\bibinfo{series}{SIGSPATIAL '19})}.
  \bibinfo{publisher}{Association for Computing Machinery},
  \bibinfo{address}{New York, NY, USA}, \bibinfo{pages}{139–148}.
\newblock
\showISBNx{9781450369091}
\urldef\tempurl%
\url{https://doi.org/10.1145/3347146.3359077}
\showDOI{\tempurl}


\bibitem[\protect\citeauthoryear{Zhang, Lim, Leis, Andersen, Kaminsky, Keeton,
  and Pavlo}{Zhang et~al\mbox{.}}{2018}]%
        {surf}
\bibfield{author}{\bibinfo{person}{Huanchen Zhang}, \bibinfo{person}{Hyeontaek
  Lim}, \bibinfo{person}{Viktor Leis}, \bibinfo{person}{David~G. Andersen},
  \bibinfo{person}{Michael Kaminsky}, \bibinfo{person}{Kimberly Keeton}, {and}
  \bibinfo{person}{Andrew Pavlo}.} \bibinfo{year}{2018}\natexlab{}.
\newblock \showarticletitle{SuRF: Practical Range Query Filtering with Fast
  Succinct Tries} \emph{(\bibinfo{series}{SIGMOD '18})}.
  \bibinfo{publisher}{Association for Computing Machinery},
  \bibinfo{address}{New York, NY, USA}, \bibinfo{pages}{323–336}.
\newblock
\showISBNx{9781450347037}
\urldef\tempurl%
\url{https://doi.org/10.1145/3183713.3196931}
\showDOI{\tempurl}


\end{thebibliography}

\end{document}